\documentclass[]{aastex631}
\usepackage[utf8]{inputenc}
\usepackage{natbib}
\setcitestyle{round}
\usepackage{amsmath,amssymb}
\usepackage{xcolor}
\usepackage[caption=false]{subfig}
\usepackage{soul}
\usepackage{enumitem}
\usepackage{xcolor}
\usepackage{comment}
\usepackage{gensymb}
\usepackage{slantsc}
\usepackage{lmodern}

\begin{document}

\title{Metallicity Structure in Galactic Longitude-Velocity Diagrams of the Milky Way Disk and FIRE-2 Simulations}

\correspondingauthor{Victor Liu}

\author[0000-0003-4434-3921]{Victor Liu}
\altaffiliation{Department of Astronomy \& Astrophysics, Pennsylvania State University, 525 Davey Laboratory, University Park, PA 16802, USA}
\affiliation{National Radio Astronomy Observatory, 520 Edgemont Road, Charlottesville, VA 22903, USA}
\email{victorliu1231@gmail.com}
\author[0000-0002-2465-7803]{Dana S. Balser}
\affiliation{National Radio Astronomy Observatory, 520 Edgemont Road, Charlottesville, VA 22903, USA}
\email{dbalser@nrao.edu}
\author[0000-0003-3351-6831]{Trey V. Wenger}
\affiliation{NSF Astronomy \& Astrophysics Postdoctoral Fellow, Department of Astronomy, University of Wisconsin--Madison, 475 N Charter St, Madison, WI 53703, USA}
\affiliation{California State University, Chico, 400 West First Street, Chico, CA 95929, USA}
\email{tvwenger@csuchico.edu}

\begin{abstract}
We investigate longitude-velocity ($\ell$-$v$) diagrams as a diagnostic tool to study the metallicity structure of the Milky Way (MW) disk. The present-day metallicity structure encodes the imprint of the Galaxy's formation, assembly, and secular evolution. Using oxygen abundances from H\,{\sc ii} regions across the MW disk, together with MW-mass galaxies from the Feedback in Realistic Environments (FIRE-2) cosmological simulations, we show that $\ell$-$v$ diagrams trace radial metallicity gradients and non-axisymmetric azimuthal metallicity variations. Because they do not rely on distance measurements, $\ell$-$v$ diagrams complement face-on maps for studying metallicity structure. In the MW, we detect the radial metallicity gradient in $\ell$-$v$ space, but current H\,{\sc ii} region oxygen abundance errors are too high to reveal azimuthal variations. In the FIRE-2 MW-mass galaxies, the radial gradient is evident in $\ell$-$v$ diagrams regardless of observer location, but anomalous gas kinematics can mimic azimuthal metallicity variations. We term these ``anomalous motions'', which have an excess local standard of rest (LSR) velocity tail 3 times larger in the FIRE-2 simulations compared to the MW. Our results highlight $\ell$-$v$ diagrams as a largely unexplored tool for probing metallicity structure without requiring distances, and underscore discrepancies between the gas kinematics in the FIRE-2 simulations and those in the MW.
\end{abstract}

\section{Introduction}
The detailed chemical structure of the Milky Way (MW) disk offers key insights into the MW's formation history, star formation efficiency, and the dynamical processes critical in the MW's evolution. Studies reveal a negative radial metallicity gradient in both the gas and stars in the disks of most galaxies, including the MW (e.g., \citealt{Searle_1971, Shaver_1983, Bovy_2014, Hayden_2014}). The magnitude of these gradients depends on multiple physical processes important in Galactic evolution, such as star formation, stellar migration, radial gas inflows, and stellar feedback (for a comprehensive review see \citealt{Maiolino_Mannucci_2019}). These radial metallicity gradients are consistent with the inside-out growth theory of galaxies \citep{Fall_1980}, where the inner disk formed earlier than the outer disk and caused a higher star formation rate in the central regions early on (e.g., \citealt{Chiappini_2001, Dave_2011, Tissera_2021, Jia_2024}). The higher star formation rate in the inner disk, however, can also be produced from the combination of gas density declining with radius and a super-linear Kennicutt-Schmidt law (e.g. \citealt{Kennicutt_1998}) without the need for invoking inside-out growth \citep{Lacey_1985, Graf_2024}. Studies have also found that the radial metallicity gradients in the MW and simulated MW-mass galaxies are flatter with increasing distance from the disk midplane \citep{Jia_2018, Graf_2024}. Stars in the thick disk come from an earlier era with higher gas turbulence and thus have more radial mixing that wash out any radial metallicity gradients, whereas stars in the thin disk are born from gas with lower turbulence and less mixing \citep{Graf_2024}. Additionally, the inside-out scenario predicts that the radial gradient flattens over time: star formation proceeds longer in the outer disk, continually enriching the ISM at large radii and allowing the outer disk metallicity to ``catch up'' to that of the inner disk. The picture is complicated, however, and there are mixed results from simulations on whether inside-out growth necessarily causes the flattening of radial metallicity gradients over time (e.g., \citealt{Vincenzo_2018, Sharda_2021, Bellardini_etal_2021, Renaud_2025, Graf_2024}).

Constraining chemical evolution models is key to advancing our understanding of the MW's evolution. Early chemodynamical models typically assumed axisymmetric metallicity gradients (e.g., \citealt{Chiappini_2003, Schonrich_2009}), as efficient mixing was thought to erase azimuthal inhomogeneities. More recent studies, though, found non-axisymmetric azimuthal variations in both the MW \citep{Davies_2009, Balser_2015, Wenger_2019_VLA_sample} and other galaxies \citep{Sanchez-Menguiano_2016, Sanchez-Menguiano_2017, Ho_2017, Ho_2018, Kreckel_2019, Gillman_2022, Neumann_2024, Chen_2024}. In the MW, these azimuthal abundance variations are detected from electron temperatures in H\,{\sc ii} regions \citep{Balser_2015, Wenger_2019_VLA_sample} and iron abundances in Cepheids \citep{Luck_2006, Pedicelli_2009}. Several theories have been proposed to explain these variations: (1) large-scale non-circular gas motions driven by galactic spiral arms and/or bars for gas-phase azimuthal variations \citep{Orr_etal_2023_metalfreeways}; (2) stellar radial migration induced by spiral arm perturbations for azimuthal variations in young stars \citep{Grand_etal_2016}; and (3) the trapping of young, metal-rich stars in the gravitational potential of spiral arms for azimuthal variations across all stars \citep{Khoperskov_2018}. Each theory predicts azimuthal metallicity variations of comparable magnitude to those observed. Distinguishing between these theories will offer insight into the impact of spiral arms, the Galactic bar, and large-scale streaming motions on chemical evolution.

H\,{\sc ii} regions are excellent probes of chemical structure in the MW for several reasons. (1) They are bright and therefore detected throughout the MW disk \citep{Wenger_2021_SHRDS_II}; (2) H\,{\sc ii} regions are young with lifetimes less than $\sim$10 Myr \citep{Yorke_1986_HII_region_theory}, so their abundances correspond to the present day chemical structure; and (3) several methods are available to estimate H\,{\sc ii} region distances. Traditionally, collisionally excited lines (CELs) in the optical and infrared are used to probe the abundances of relevant species like oxygen and iron, as these lines are directly produced from the different ionization states of these metals (e.g., \citealt{Aller_1984, Bresolin_2009, Perez-Montero_2014}). 

Radio recombination lines (RRLs) and free-free radio continuum emission are an alternative tracer of Galactic H\,{\sc ii} regions. Unlike optical and infrared CELs, radio tracers are largely unaffected by interstellar dust, enabling the detection of thousands of previously obscured nebulae (see \citealt{Wenger_2019_ATCA_RRL} for a brief history of H\,{\sc ii} region RRL surveys). The RRL-to-continuum intensity ratio is an accurate measure of the electron temperature of an H\,{\sc ii} region assuming local thermodynamic equilibrium (LTE) and an optically thin nebula (e.g., \citealt{Mezger_1967, Wenger_2019_VLA_sample}). Since metals primarily regulate the temperature in H\,{\sc ii} regions, the electron temperature is a proxy for metallicity \citep{Rubin_1985}. Metals are effective coolants for gas because they have low-lying excited states just a few eV above ground that can easily be excited by collisions with electrons, and metals have a high cooling efficiency per atom. \citet{Shaver_1983} empirically showed that the O/H abundance ratio is inversely proportional to the electron temperature for a sample of H\,{\sc ii} regions. \citet{Balser_2024} further refined the metallicity-electron temperature relationship to account for non-LTE effects, variations in the electron density, and the spectral type of the ionizing star.

Constructing the spatial distribution of H\,{\sc ii} region metallicities requires distances. Kinematic distance determinations are subject to two primary challenges (see \citealt{Burton_1971} for a comprehensive review). First, within the solar orbit, any given Galactic longitude and local standard of rest velocity ($v_{\rm LSR}$) correspond to two possible spatial locations, a degeneracy known as the kinematic distance ambiguity (KDA). If not properly resolved, the KDA can introduce large systematic uncertainties in distances. Second, near the terminal velocity inside the solar orbit and at the cardinal Galactic directions ($\ell=0\degree,90\degree,180\degree,$ and $270\degree$), emission at slightly different velocities may originate from significantly different distances, an effect known as velocity crowding. Due to these issues, mapping emission using distances is prone to misinterpretation (see \citealt{Burton_1971, Burton_1974, Peek_2022}). Although parallax measurements have begun to improve distance constraints (e.g., \citealt{Reid_2014, Reid_2019, Wenger_2018, Andriantsaralaza_2022, Simon_2024}), we choose to use only longitude-velocity ($\ell$-$v$) diagrams, which are distance-independent, to study metallicity structure in the MW and to avoid complications with kinematic distances.

Cosmological simulations are powerful tools for studying Galactic metallicity structure because they provide fully resolved spatial, kinematic, and chemical information, allowing direct comparisons between metallicity maps in $\ell$-$v$ and $x$-$y$ space. The Feedback in Realistic Environments (FIRE-2) simulations are a suite of publicly available cosmological zoom-in simulations that focus on accurately modeling stellar feedback in galactic evolution \citep{Hopkins_etal_2018_FIRE}. The FIRE-2 Latte and ELVIS suites of MW-mass galaxy simulations have been used extensively to study properties of MW and MW-like galaxies, such as their chemical evolution \citep{Bellardini_etal_2021, Orr_etal_2023_metalfreeways, Graf_2024}, drivers of star formation \citep{Orr_etal_2018_KS_relation, Orr_etal_2020}, and bar properties and formation histories \citep{Debattista_etal_2019, Ansar_2025}. Multiple studies have detected and explored both radial and azimuthal metallicity variations in the Latte and ELVIS suite (e.g., \citealt{Bellardini_etal_2021, Orr_etal_2023_metalfreeways, Graf_2024}). 

We explore metallicity structure in $\ell$-$v$ diagrams for both the MW and simulated MW-mass galaxies from the FIRE-2 suite. Unlike traditional face-on maps, which rely on distance estimates often uncertain in the MW, $\ell$-$v$ diagrams use directly observed kinematic and positional information, making them uniquely suited to reveal large-scale metallicity features such as radial gradients and azimuthal variations. By applying this diagnostic to both observations and simulations, we aim to assess the $\ell$-$v$ diagram's potential to probe the present-day chemical structure of the Galactic disk.

\section{Data}
\subsection{Milky Way H\,\textsl{\textsc{ii}} Regions} \label{subsec:MW_data}
The census of Galactic H\,{\sc ii} regions has more than doubled over the past decade and a half because of the H\,{\sc ii} Region Discovery Surveys (HRDSs; \citealt{Bania_2010_GBT_survey_I, Bania_2012, Anderson_2011_GBT_survey_II, Anderson_2015, Anderson_2018, Wenger_2019_ATCA_RRL, Wenger_2021_SHRDS_II}). These RRL surveys unambiguously identify the thermal emission from the H\,{\sc ii} regions and these nebulae are included in the WISE Catalog of Galactic H\,{\sc ii} regions\footnote{The WISE Catalog v2 is at \url{https://doi.org/10.7910/DVN/NQVFLE} \citep{Wenger_2021_WISE_catalog} and online at \url{http://astro.phys.wvu.edu/wise/}.} as ``known'' H\,{\sc ii} regions (\citealt{Anderson_2014}; hereafter the WISE Catalog). Included in the WISE Catalog are $\sim$800 H\,{\sc ii} regions with accurate electron temperatures, in part based on more sensitive and better calibrated RRL and continuum data \citep{Quireza_2006_RRL, Balser_2011, Balser_2015, Wenger_2019_VLA_sample}. To determine the oxygen abundance we use these electron temperatures together with the following empirical relationship from \citet{Shaver_1983}
    \begin{equation} \label{eq:Te_to_OH}
        12 + {\rm log}_{10}({\rm O/H}) = (9.82\pm0.02) - (1.49\pm0.11) \frac{T_{\rm e}}{10^4 {\rm K}},
    \end{equation}
where ${\rm O/H}$ is the oxygen abundance ratio by number and $T_{\rm e}$ is the electron temperature.

In reality, the metallicities and electron temperatures of H\,{\sc ii} regions slightly deviate from this relationship, which assumes the H\,{\sc ii} region is in pure LTE. Non-LTE effects in H\,{\sc ii} regions necessitate a re-calculation of the electron temperature from the RRLs and radio continuum, changing the slope and offset of Equation \ref{eq:Te_to_OH}. Using Cloudy simulations\footnote{Cloudy is a spectral synthesis code that models the physical conditions and emission spectra of gas clouds \citep{CLOUDY_2023}.}, \citet{Balser_2024} quantifies the shift in the relationship due to these non-LTE effects and supplies a list of oxygen abundance correction factors as a function of the metallicity of the H\,{\sc ii} region, the electron density, and the spectral type of the ionizing star. Higher values of the electron density will increase the rate of collisional de-excitation between electrons and ionized metals, thus inhibiting cooling and slightly increasing $T_{\rm e}$ \citep{Rubin_1985}. A hotter stellar effective temperature of the ionizing star increases the hardness of the radiation field that excites and heats the gas, causing a marginal increase in $T_{\rm e}$ \citep{Rubin_1985}. Regardless, the electron temperature is an important tracer of metallicity as most Galactic H\,{\sc ii} regions only deviate slightly from LTE \citep{Shaver_1980}. 

To estimate the oxygen abundance errors, we add the observed oxygen abundance errors, $\sigma_{\rm obs}$, and the systematic errors, $\sigma_{\rm sys}$, in quadrature. To derive $\sigma_{\rm obs}$, we propagate the WISE Catalog electron temperature errors, which are 1$\sigma$ uncertainties derived from the continuum and RRL uncertainties, through Equation \ref{eq:Te_to_OH}. We don't include Equation \ref{eq:Te_to_OH}'s error in the intercept and slope in this error propagation since these are systematic errors induced by fitting to a spread of H\,{\sc ii} regions with different properties, such as the electron density and the spectral type of the ionizing star. The effect of this spread in H\,{\sc ii} region properties on the oxygen abundance ratio calculation is already captured by the correction factors (CFs) from \citet{Balser_2024} and causes $\sigma_{\rm sys}$. We use these CFs to estimate $\sigma_{\rm sys}$. The subsample of H\,{\sc ii} regions in the WISE catalog that contain electron temperatures are typically ``classical'' H\,{\sc ii} regions, i.e., neither ultra-compact nor ultra-diffuse \citep{Quireza_2006_RRL, Balser_2011, Balser_2015, Wenger_2019_VLA_sample}. From \citet{Balser_2024}, classical H\,{\sc ii} regions with typical metallicities correspond to a range of CFs between 0.97 and 1.03. The fractional error equals the range of these CFs, or 0.06, so the systematic error is $\sigma_{\rm sys}=0.06*{\rm log_{10}(O/H)}$. We use the CFs only to estimate $\sigma_{\rm sys}$. In principle we could apply the correction factor to the oxygen abundances instead by estimating the nebular electron density and spectral type of the ionizing star. This requires distances, however, which we avoid invoking in this paper. The systematic errors dominate over the observed errors, with the error ratio, $\sigma_{\rm sys}/\sigma_{\rm obs}$, typically equal to $\sim$4.78.

\begin{figure}[h]
        \centering
        \includegraphics[width=1\linewidth]{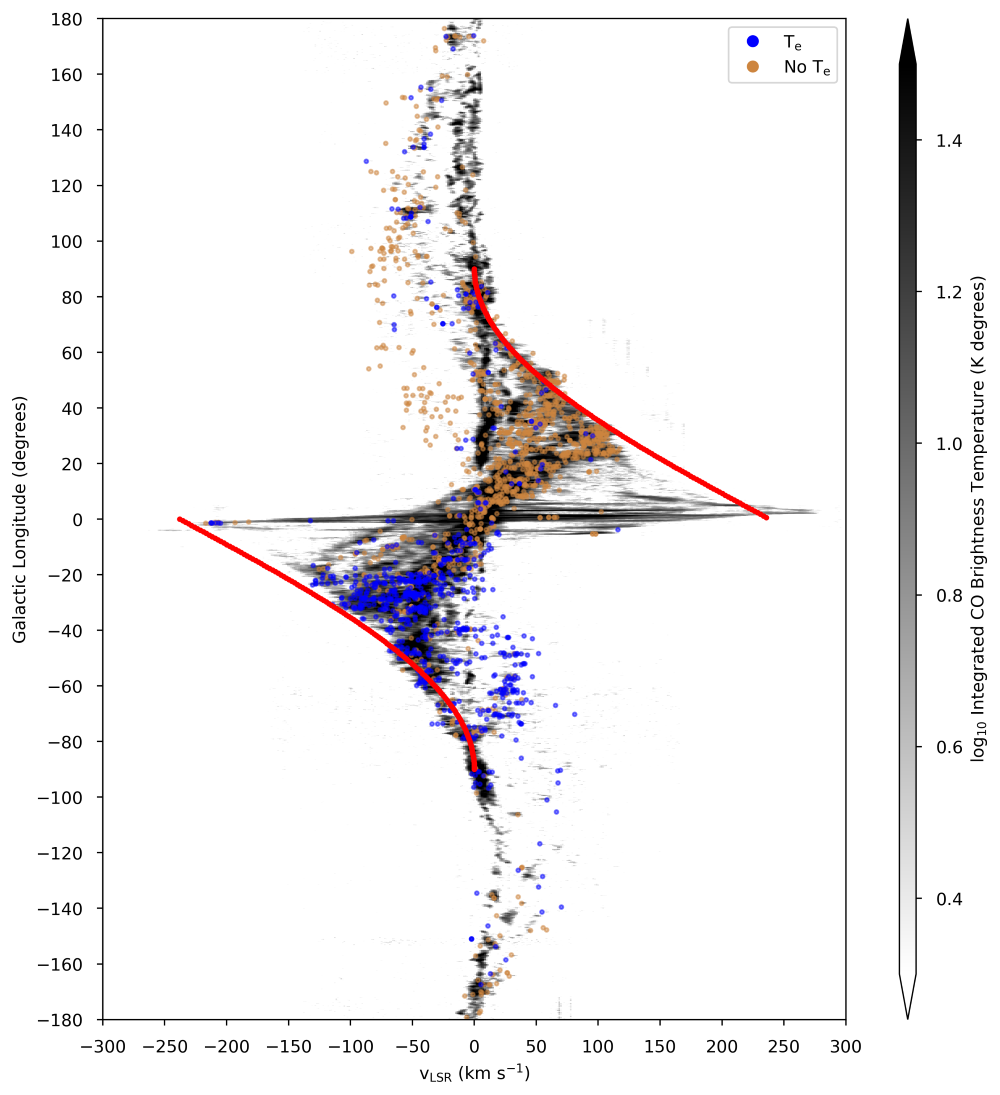}
        \caption{The $\ell$-$v$ diagram for all known MW H\,{\sc ii} regions from the WISE Catalog. There are 737 H\,{\sc ii} regions with derived electron temperatures and 1295 H\,{\sc ii} regions with no derived electron temperature, colored blue and brown respectively. Of the 737 sources with electron temperatures, 686 have accurately determined electron temperatures (see text). There are significantly more H\,{\sc ii} regions with derived electron temperatures in the Galactic Southern Hemisphere compared to the Northern Hemisphere. Superimposed is the CO(J=1$\rightarrow$0) brightness temperature from \citet{Dame_2001_CO} integrated over $|b| \leq3\degree$. The CO emission has a typical rms of 1.15 K degrees, or 0.06 K degrees in ${\rm log}_{10}$ units. The color bar range is 0.3$-$1.5 K degrees in ${\rm log}_{10}$ units. The red curve is the terminal velocity for the MW using $R_{\rm 0,MW}$ = 8.20 kpc for the solar Galactocentric radius and $\Theta_{\rm 0,MW}$ = 238 km $\rm s^{-1}$ for the circular rotation speed at the Sun's position.}
        \label{fig:MW_LV}
    \end{figure}

Figure \ref{fig:MW_LV} shows the $\ell$-$v$ diagram of known H\,{\sc ii} regions in the WISE Catalog. Overplotted is the distribution of CO(J=1$\rightarrow$0) emission from \citealt{Dame_2001_CO} as a proxy for star-forming cold molecular clouds. Sources with derived electron temperatures are colored blue and those without are colored brown. The terminal velocity is the maximum line-of-sight velocity expected from purely circular motion along a given line of sight within the solar orbit. The red curve represents the MW's terminal velocity, calculated using a solar Galactocentric radius of $R_{\rm 0,MW} = 8.20$ kpc and a circular rotation speed at the Sun's location of $\Theta_{\rm 0,MW} = 238$ km $\rm {s}^{-1}$. These parameters are broadly consistent with modern studies (see \citealt{Leung_2022} and \citealt{Bland-Hawthorn_2016} for relevant reviews). For example, \citet{Reid_2019} found $R_{\rm 0,MW} = 8.15\pm0.15$ kpc and $\Theta_{\rm 0,MW} = 236\pm7$ km $\rm {s}^{-1}$. Some emission---both H\,{\sc ii} regions and CO---deviate from the expected velocities of the flat rotation curve, reflecting the presence of non-circular motions and variations in rotation speed driven by non-axisymmetric structures like the Galactic bar and spiral arms.

In order to study Galactic metallicity structure, accurate electron temperatures are needed. In the WISE Catalog, only 737 known sources have derived electron temperatures. These electron temperatures are computed from the RRL-to-continuum intensity ratio. The quality of the continuum and spectral line measurements for most of these sources are ranked from 1 to 5 via visual inspection, where 1 corresponds to the best quality and 5 to the worst. We exclude sources with continuum and/or spectral line quality factors of 4 or 5, resulting in a final sample of 686 sources (88 in the Northern sky and 598 in the Southern sky). Among these, 82 lack derived observational electron temperature errors. Nevertheless, since the systematic error dominates over the observational error, we still retain these sources in the dataset.

\subsection{FIRE-2 Simulations}
For the FIRE-2 simulations, we use the following Latte suite galaxies: m12i \citep{Wetzel_2016_m12i}, m12f \citep{Garrison-Kimmel_2017_m12f_RomeoJuliet_ThelmaLouise}, and m12m \citep{Hopkins_etal_2018_FIRE}. The Latte galaxies evolve in an isolated halo environment. From the ELVIS suite, we use Romeo \citep{Garrison-Kimmel_2017_m12f_RomeoJuliet_ThelmaLouise} and Romulus and Remus \citep{Garrison-Kimmel_2019_RomulusRemus}. Unlike the Latte suite, the ELVIS suite galaxies evolve in a Local Group (LG)-like environment with two gravitationally interacting halos similar to the Milky Way + Andromeda pair. For all galaxies, we only analyze the snapshot at z=0. All of these galaxies were simulated using the GIZMO code \citep{Hopkins_2015_GIZMO}.\footnote{All of these simulations are publicly available at \url{https://flathub.flatironinstitute.org/fire} and fully described by \citet{Wetzel_2023_FIRE2DataRelease}.}

Within the Latte suite, m12i, m12f, and m12m are most like the MW, having broadly similar stellar masses, scale radii, scale heights, and gas fractions \citep{Sanderson_2020}. In the ELVIS suite, Romeo, Romulus, and Remus are most similar to the MW, having the thinnest, most MW-like disks \citep{Wetzel_2023_FIRE2DataRelease}. Compared to the Latte suite, the galaxies in the ELVIS suite tend to have larger disks \citep{Bellardini_2022} and begin forming and settling their disks earlier \citep{Santistevan_2020, Yu_2021, McCluskey_2023}. We use Romeo as our benchmark for comparing the FIRE-2 simulations to the MW since Romeo contains the earliest forming disk \citep{Wetzel_2023_FIRE2DataRelease} and the MW is thought to exhibit early disk formation (e.g., \citealt{Belokurov_2022, Conroy_2024}). Table \ref{tab:sim_properties} summarizes the z = 0 properties of the FIRE-2 simulations and relevant quantities for our $\ell$-$v$ diagram analysis. Listed are the virial mass, $M_{\rm 200m}$, the virial radius, $R_{\rm 200m}$, the enclosed stellar mass, $M_{\rm star,90}$, the galactocentric radius of the observer, $R_{obs}$, and the mean rotational velocity of cold, dense, and neutral (C\&D) gas, $\bar{V}_{\rm rot,C\&D}$. The virial mass and virial radius, $M_{\rm 200m}$ and $R_{\rm 200m}$, are the total mass and spherical radius within which the mean density is 200 times the matter density of the universe, respectively. The enclosed stellar mass, $M_{\rm star,90}$, is the stellar mass within a spherical radius that encloses 90\% of the stellar mass within 20 kpc. We place the simulated observer in these galaxies at $R_{\rm obs}=(\frac{R_{\rm 0,MW}}{R_{\rm 200m,MW}})R_{\rm 200m}$ in order to mimic the Sun's position within the Milky Way. We use $R_{\rm 0,MW} = 8.20$ kpc, a value in rough agreement with most modern studies of $R_{\rm 0,MW}$ (see Figure 1 of \citealt{Leung_2022}). For the virial radius of the MW, $R_{\rm 200m,MW}$, we use 220 kpc, which is an intermediate value between the cited values in the literature (see Table 8 of \citealt{Bland-Hawthorn_2016}). The cited values for $R_{\rm 0,MW}$ and $R_{\rm 200m,MW}$ vary in the literature and are constantly being updated as new data and techniques become available. To be robust, we test whether varying $R_{\rm obs}$ has an effect on the detection of metallicity trends observed in $\ell$-$v$ space (see Section \ref{sec:results}). 

\begin{deluxetable}{lccccc}[h]
    \tablecaption{FIRE-2 MW-like galaxies z=0 properties}
    \tablehead{\colhead{Simulation}  & \colhead{$M_{\rm 200m}$}  & \colhead{$R_{\rm 200m}$} & \colhead{$M_{\rm star,90}$} & \colhead{$R_{\rm obs}$} & \colhead{$\bar{V}_{\rm rot,C\&D}$} \\
         Name & $(M_\odot)$ & (kpc) & $(M_\odot)$ & (kpc) & (km ${\rm s}^{-1}$)}
    \startdata
        m12i & 1.18e12     & 336        & 6.3e10 & 12.52 & -227.52 \\ 
        m12f & 1.71e12     & 380        & 7.9e10 & 14.16 & -259.38 \\
        m12m & 1.58e12     & 371        & 1.1e11 & 13.83 & -290.89 \\ \hline 
        Romeo & 1.32e12    & 341        & 6.6e10 & 12.71 & -243.28 \\ 
        Romulus & 2.08e12 & 406         & 9.1e10 & 15.13 & -254.65 \\
        Remus & 1.22e12   & 339         & 4.6e10 & 12.64 & -218.88 \\
    \enddata
    \tablecomments{The virial mass, $M_{\rm 200m}$, and virial radius, $R_{\rm 200m}$, are defined as the total mass and spherical radius within which the mean density is 200 times the matter density of the universe, respectively. The enclosed stellar mass, $M_{\rm star,90}$, is the stellar mass within a spherical radius that encloses 90\% of the stellar mass within 20 kpc. Values are taken from \citet{Wetzel_2023_FIRE2DataRelease}. The observer galactocentricradius, $R_{\rm obs}$, is the simulated observer's galactocentric radius. The rotational velocity, $\bar{V}_{\rm rot,C\&D}$, is the mean rotational velocity of cold, dense, and neutral gas in the simulated galaxy (see text).
    }
    \label{tab:sim_properties}
\end{deluxetable}

These simulated galaxies represent one of the highest mass and spatial resolutions for zoom-in cosmological simulations to date, making them well-suited for studying the spatial distribution of chemical abundances. The Latte suite galaxies have a gas cell mass resolution of $m_{\rm gas} = 7100$ $\rm M_{\odot}$ and an initial star particle mass resolution of $m_{\rm 0,star} = 7100$ $\rm M_{\odot}$. The ELVIS suite galaxies have mass resolutions of $m_{\rm gas} = 3500$ $\rm M_{\odot}$ and $m_{\rm 0,star} = 3500$ $\rm M_{\odot}$. Star particles are treated as single stellar populations, with a singular age, metallicity, and mass. The metallicity and mass are inherited from the progenitor gas cell, assuming a \citet{Kroupa_2001} initial mass function. The star particle's mass resolution typically decreases by $\sim$30\% over time because of stellar mass loss. Both suites have a minimum gas cell spatial resolution $\leq$ 1.0 pc and a fixed star particle spatial resolution $=$ 4.0 pc. The gas cell spatial resolution is adaptive and automatically adjusts to the local gas density and physical conditions.

The Latte and ELVIS suite simulated galaxies have virial masses of $1-2 \times 10^{12}$ $\rm M_\odot$, similar to the Milky Way \citep{Kafle_2012, Watkins_2019}. The major and minor axes of the galactic plane are defined by the eigenvectors of the moment-of-inertia tensor of young star particles, which is computed by integrating the stellar mass density distribution over the volume of the galaxy. A similar scheme is used to define the orientation of the disks in the Auriga zoom-in cosmological simulations \citep{Grand_etal_2017}. The FIRE-2 simulations in this study do not include supermassive black holes (SMBHs) and do not model SMBH feedback. For a full description of the FIRE-2 simulation initial conditions, physics, and galaxy properties, see \citet{Hopkins_etal_2018_FIRE} and \citet{Wetzel_2023_FIRE2DataRelease}.

The FIRE-2 simulations model metallicities by including an explicit sub-grid model for the turbulent diffusion of elemental abundances in gas, which accounts for mixing driven by unresolved eddies and effectively smooths out abundance differences between neighboring gas elements \citep{Su_2017, Escala_2017, Hopkins_etal_2018_FIRE}. Including this sub-grid mixing is essential to reproduce the observed spread in stellar metallicities \citep{Escala_2017}. While the details of this diffusion model do not significantly impact the radial or vertical metallicity gradients in FIRE-2 MW-mass galaxies, they do affect the azimuthal scatter at a fixed radius \citep{Bellardini_etal_2021}. The stellar metallicity radial gradients in these simulated galaxies also tend to be shallower than those measured in the MW \citep{Bellardini_2022, Graf_2024}. Regardless, the FIRE-2 MW-mass galaxies exhibit radial gradients and azimuthal metallicity variations in the ISM that are consistent with those observed in nearby galaxies of similar mass \citep{Bellardini_etal_2021}.

To probe the C\&D gas in the galaxy, we select gas cells with temperature $T < 500$ K, number density $n_{\rm H} > 1$ ${\rm  cm}^{-3}$, and ionized hydrogen mass fraction identically equal to 0. Following the selection criteria of \citet{Orr_etal_2023_metalfreeways}, the C\&D gas is taken as a proxy for cold molecular gas traced by cold dust or CO observations. Here, we select C\&D gas cells that are located $|x| \leq$ 20 kpc, $|y| \leq$ 20 kpc, and $|z| \leq$ 7.5 kpc. We also select young stellar populations (YSPs) which are star particles with ages $<$ 10 Myr as a tracer for H\,{\sc ii} regions. The 10 Myr time interval was chosen for its approximate correspondence with timescales in star-forming clouds traced by recombination lines like H$\alpha$ \citep{Kennicutt_2012}. Spatially, we cut the YSPs on $|x| \leq$ 20 kpc, $|y| \leq$ 20 kpc, and $|z| \leq$ 2.5 kpc. We use different $|z|$ cutoffs since almost all of the C\&D gas is located at $|z|\leq$ 7.5 kpc, whereas almost all of the YSPs are located at $|z|\leq$ 2.5 kpc. As our observational samples encompass the known CO emission and H\,{\sc ii} regions within the MW, we choose these spatial ranges in the simulations to likewise encapsulate all the C\&D gas and YSPs.

FIRE-2 tracks 11 elements (H, He, C, N, O, Ne, Mg, Si, S, Ca, Fe). Of these 11 elements, we choose oxygen as our metallicity tracer to compare the metallicities between the C\&D gas and YSPs in the FIRE-2 galaxies and the H\,{\sc ii} regions in the MW. Note that studies cataloging the oxygen abundances in MW H\,{\sc ii} regions typically express the oxygen abundance as the ratio of oxygen to hydrogen by number density. In contrast, FIRE-2 expresses the oxygen abundances in terms of mass fractions (i.e., O/H is the ratio of the mass fraction of oxygen relative to hydrogen). Therefore, we convert the FIRE-2 oxygen abundances into number densities using
    \begin{equation} 
    \label{eq:massfrac_to_num_density}
        {\rm O/H} \equiv \frac{n_{\rm O}}{n_{\rm H}}  = \frac{w_{\rm O}/m_{\rm O}}{w_{\rm H}/m_{\rm H}},
    \end{equation}
where \textit{n} is the number density, \textit{w} is the mass fraction of the element, and \textit{m} is the atomic mass of the element. We use atomic masses $m_{\rm O} = 15.999$ amu and $m_{\rm H} = 1.0080$ amu \citep{Prohaska_2022_atomic_weights}. 

\section{Results} \label{sec:results}
\begin{figure}
    \centering
    \includegraphics[width=1\linewidth]{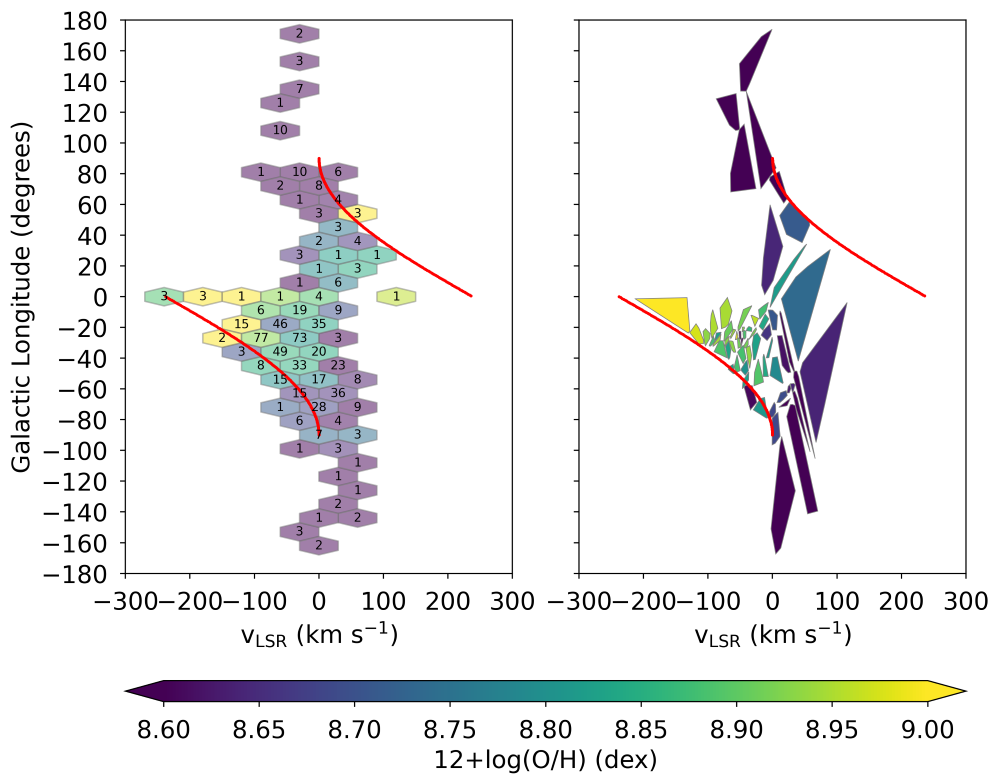}
    \caption{$\ell$-$v$ diagrams for known H\,{\sc ii} regions in the Milky Way from the WISE Catalog, colored by oxygen abundance. The two panels display different binning schemes. The left panel bins the points into equally sized hexagons labeled by the number of points inside. The right panel uses the Voronoi tessellation technique to bin the points such that each bin roughly has an equal number of points (see text). In the right panel, the boundary of each polygon connects the outer points in that bin, and each bin contains between 11 and 14 points. Both panels plot the mean oxygen abundance in each bin. The red line is the terminal velocity curve for the MW (described in the caption of Figure \ref{fig:MW_LV}). In both visualization schemes, the negative radial metallicity gradient is visible when moving from low $|\ell|$ to high $|\ell|$ along the terminal velocity curve, and also when moving from high $|v_{\rm LSR}|$ to low $|v_{\rm LSR}|$ at a fixed Galactic longitude.
    }
    \label{fig:MW_abundances}
\end{figure}

Our goal is to characterize metallicity structure, specifically the radial gradient and azimuthal variations, in $\ell$-$v$ diagrams. Figure \ref{fig:MW_abundances} shows two $\ell$-$v$ diagrams of the metallicity of known MW H\,{\sc ii} regions from the WISE Catalog. In each diagram, the regions are binned and colored by the mean oxygen abundance. In the left panel, the bins are shaped as equally sized hexagons and each bin is labeled by the number of points inside. The bins in the Northern hemisphere and on the edges of the $\ell$-$v$ diagram in the Southern hemisphere typically contain $<5$ points. Because of small number statistics in these bins, we also use a Voronoi tessellation-like binning scheme, shown in the right panel. This tessellation technique bins 2D space into convex polygons to ensure uniformity within each bin (see \citealt{Cappellari_2003}). In \citet{Cappellari_2003}, the tessellations are used to reach a constant signal-to-noise ratio per bin for their integral-field spectroscopic (IFS) data, but here we use the tessellations to reach a similar number of points per bin for our H\,{\sc ii} region data. We use a kd-tree to implement the Voronoi tessellation technique. This allows us to compare mean metallicities between bins with similar statistics. See Appendix \ref{sec:appendix_equations} for details about generating $\ell$-$v$ diagrams for the FIRE-2 simulations.

\begin{figure}[h]
    \centering
    \includegraphics[width=1\linewidth]{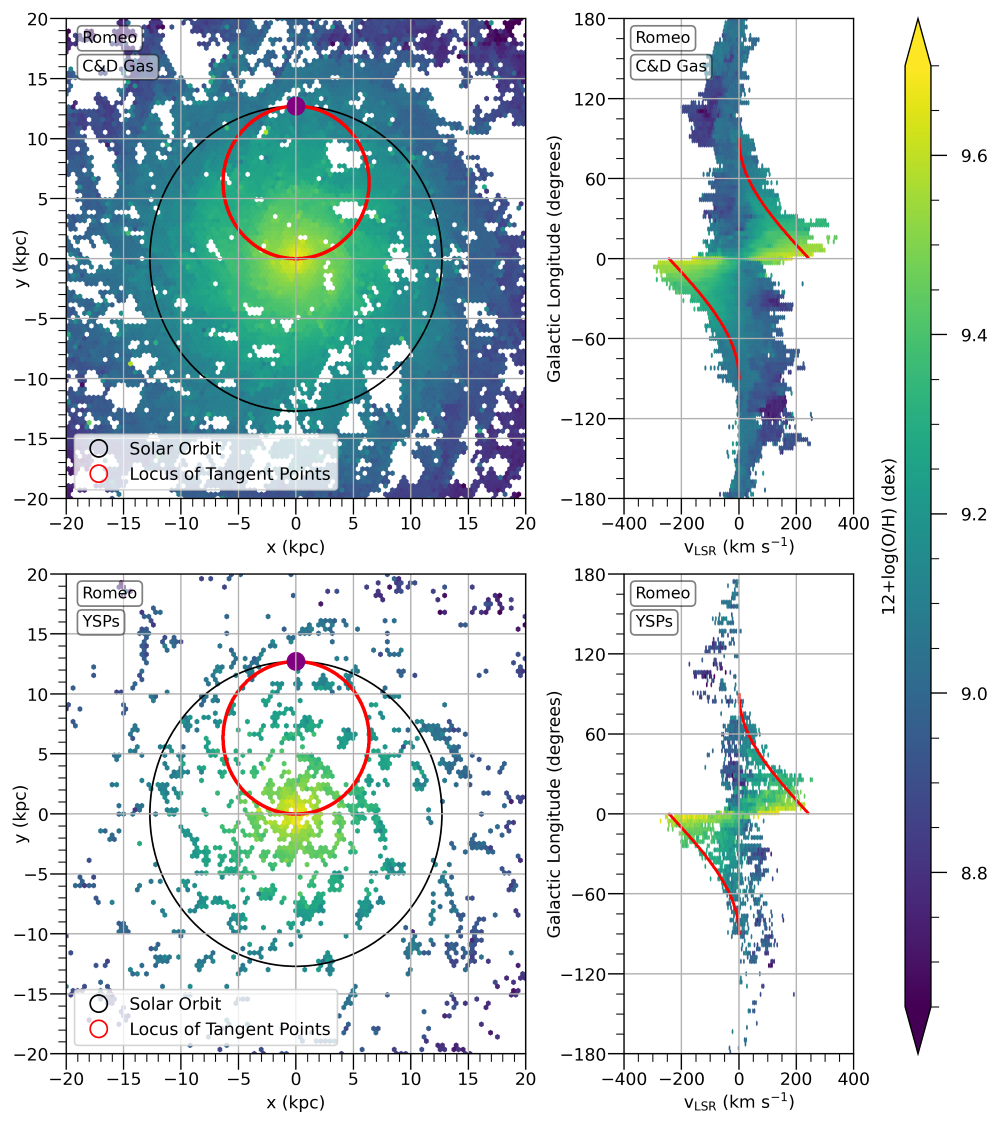}
    \caption{Oxygen abundances in the FIRE-2 Romeo galaxy at z=0. \textit{Left column:} face-on maps of Romeo. The black circle is the solar orbit and the red circle is the locus of tangent points. Here, solar orbit refers to points with the same galactocentric radius as the observer. The observer is located at $(R,\theta)$ = $(12.71$ kpc, $90\degree)$, as indicated by the purple dot. \textit{Right column:} corresponding $\ell$-$v$ diagrams. The red curve is the terminal velocity, which corresponds to the locus of tangent points in the face-on maps in the left column. \textit{Top row:} cold and dense and neutral (C\&D) gas, which have temperature $T < 500$ K, number density $n_{\rm H} > 1$ ${\rm  cm}^{-3}$, and ionized hydrogen mass fraction = 0. \textit{Bottom row:} young stellar populations (YSPs), which have age $<$ 10 Myr. We only examine C\&D gas and YSPs with $|x| \leq$ 20 kpc and $|y| \leq$ 20 kpc. For the C\&D gas, we further cut on $|z| \leq$ 7.5 kpc. For the YSPs, the vertical height cut is $|z| \leq$ 2.5 kpc to restrict the YSPs to the disk midplane. For all four plots, we bin the oxygen abundances into hexagonal-shaped bins and plot the mean oxygen abundance within each bin. Note the clear presence of the negative radial metallicity gradient when moving from low $|\ell|$ to high $|\ell|$ along the terminal velocity curve, and when moving from high $|v_{\rm LSR}|$ to low $|v_{\rm LSR}|$ at a fixed galactic longitude in the $\ell$-$v$ diagrams.}
    \label{fig:Romeo_OH_abundances}
\end{figure}

There are two noticeable trends in the $\ell$-$v$ diagrams of Figure \ref{fig:MW_abundances}: the oxygen abundance decreases when (1) moving from $\ell = 0\degree$ to $|\ell|=90\degree$ along the terminal velocity curve and when (2) moving from high $|v_{\rm LSR}|$ to low $|v_{\rm LSR}|$ at a constant Galactic longitude. Figure \ref{fig:Romeo_OH_abundances} shows that these same trends are found in both the C\&D gas and the YSPs in the FIRE-2 Romeo galaxy. This figure shows the face-on maps of these two species and their associated $\ell$-$v$ diagrams from the Romeo simulation at z=0. The first trend arises because the terminal velocity is the maximum expected line-of-sight velocity at each Galactic longitude assuming a constant circular rotation for the Galactic rotation model (GRM). The terminal velocity points in $\ell$-$v$ space correspond to the tangent points in $x$-$y$ space, which are defined as the collection of closest points to the Galactic center along each Galactic longitude. The galactocentric radius of the tangent point is given by $R_{\rm tan}=R_0\,|{\rm sin(\ell)}|$, so an increase in $|\ell|$ corresponds to an increase in $R_{\rm tan}$. The metallicity of gas along the tangent points thus reveals the direction and relative magnitude of the radial metallicity gradient. We call this approach to tracing the radial gradient the ``terminal velocity curve'' method.

The second trend shows that oxygen abundance decreases when moving from high $|v_{\rm LSR}|$ to low $|v_{\rm LSR}|$ at a fixed Galactic longitude. This pattern arises because at any given longitude, gas with high $|v_{\rm LSR}|$ are located near the tangent points and gas with low $|v_{\rm LSR}|$ are located near the solar orbit. Therefore, as one moves from gas with high $|v_{\rm LSR}|$ to gas with low $|v_{\rm LSR}|$ along the line of sight, the $R_{\rm gal}$ increases and the metallicity decreases. We refer to this approach of tracing the radial metallicity gradient as the ``constant Galactic longitude'' method. Because velocity crowding complicates interpretation near $\ell \simeq 0\degree$ and $|\ell| \simeq 90\degree$, we recommend only applying this method at $10\degree < |\ell| < 75\degree$ where the impact of velocity crowding is minimal. We note that both trends persist when plotting the median oxygen abundance of each bin instead of the mean oxygen abundance. We find both trends are also insensitive to the value of the required minimum number of particles in each bin for both C\&D gas and YSPs.

Because of differential rotation we expect the MW disk to be well mixed, smoothing out any metallicity inhomogeneities at a given radius. Non-axisymmetric structures (e.g., spiral arms and bars) may introduce azimuthal metallicity structure, however, and be an important factor in Galactic chemical evolution. We calculate the standard deviation in O/H within bins located near zero $v_{\rm LSR}$ (i.e., near the solar orbit) and $v_{\rm LSR}\simeq v_{\rm terminal}$ (i.e., near the locus of tangent points) in the first and fourth quadrants. Assuming that all gas rotates at the same circular speed and have no radial motions, these fluctuations in O/H over different areas of the $\ell$-$v$ diagram will detect azimuthal variations.

\begin{figure}
    \centering
    \includegraphics[width=1\linewidth]{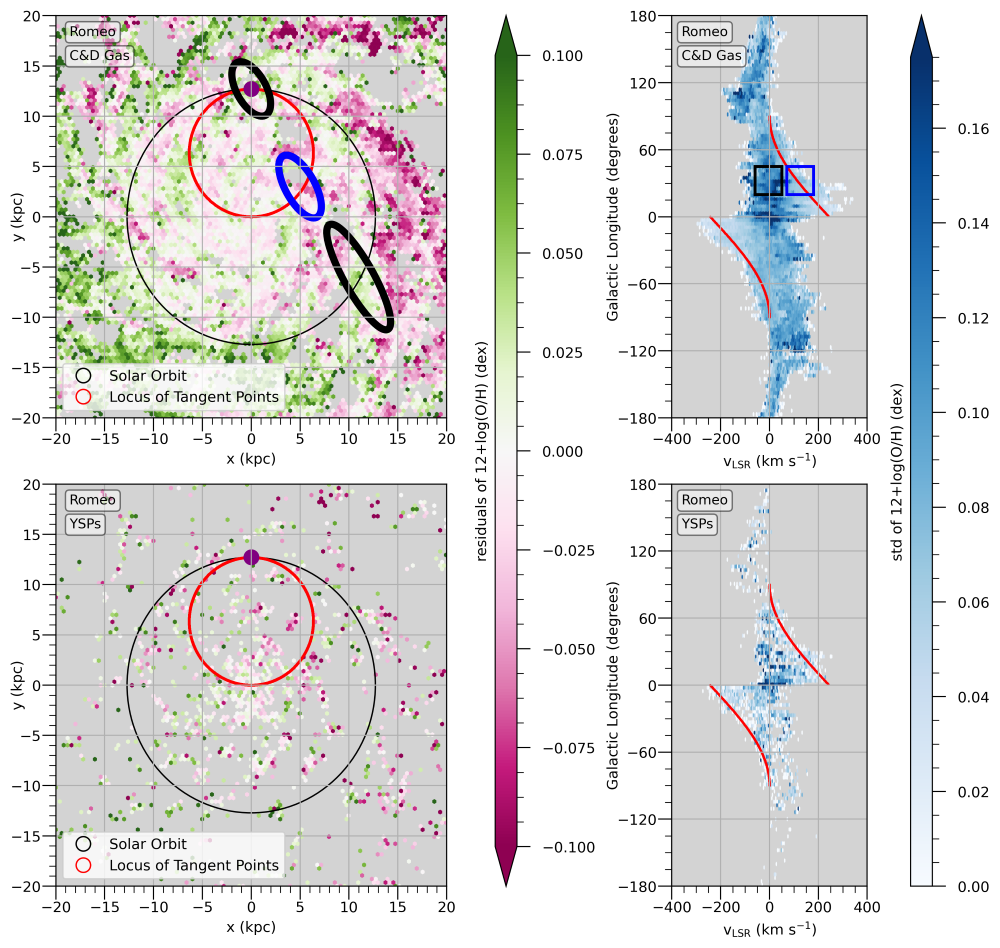}
    \caption{Azimuthal metallicity variations in the Romeo simulation. \textit{Left column}: face-on maps of Romeo colored by residual oxygen abundance, i.e., the total oxygen abundance subtracted by the radial metallicity gradient. Figure details are similar to Figure \ref{fig:Romeo_OH_abundances}. In the top-left panel, the black ellipses contain points that have $v_{\rm LSR}\simeq0$ km $\rm s^{-1}$ and $
    \ell \simeq 30\degree$, and the blue ellipse contains points that have $v_{\rm LSR}\simeq v_{\rm terminal}$ and $
    \ell \simeq 30\degree$. \textit{Right column}: corresponding $\ell$-$v$ diagrams binned in $\ell$-$v$ space and colored by the standard deviation of the oxygen abundances of cells in each bin. In the top-right panel, the black box and blue box roughly correspond to the black ellipse and blue ellipses, respectively, drawn in the top-left panel. \textit{Top row:} cold and dense and neutral (C\&D) gas. \textit{Bottom row:} young stellar populations (YSPs). See caption in Figure \ref{fig:Romeo_OH_abundances} for the definitions of C\&D gas and YSPs.Note that points in $\ell$-$v$ bins with $v_{\rm LSR} \simeq 0 \degree$ have similar $R_{\rm gal}$ but different azimuths, whereas points in $\ell$-$v$ bins with $v_{\rm LSR} \simeq v_{\rm terminal}$ typically have similar $R_{\rm gal}$ and azimuths. Anomalous motions will complicate this picture (see text). For the C\&D gas in the top-right panel, note the higher oxygen abundance scatter for bins near $v_{\rm LSR}=0$ compared to bins near the terminal velocity, indicating the presence of azimuthal metallicity variations in the C\&D gas. For the YSPs in the bottom-right panel, this feature is also present, albeit less prominently.}
    \label{fig:Romeo_OH_scatter}
\end{figure}

An example of this analysis is shown in Figure \ref{fig:Romeo_OH_scatter} using the Romeo simulation. The residual oxygen abundances are obtained by subtracting the total oxygen abundances by the radial metallicity gradient. We calculate the gradient by concatenating the mass-weighted mean metallicity for radial bins from 0 kpc to 30 kpc with a 0.1 kpc step size. Then, the residual oxygen abundance is computed by subtracting the total oxygen abundance by the mass-weighted mean metallicity of the radial bin. This procedure is done separately for the C\&D gas and YSPs. Points with high absolute LSR velocities (e.g., in the blue box in the right panel) are close to the tangent points in $x$-$y$ space (e.g., in the blue ellipse in the left panel) and thus have similar $R_{\rm gal}$ and azimuths. We expect that the metallicity scatter of these $\ell$-$v$ bins to be relatively low because the metallicity will not change significantly within these small spatial areas. In contrast, points at lower absolute LSR velocities (i.e., with $|v_{\rm LSR}| < |v_{\rm LSR,terminal}|$, like in the black box in the right panel) are azimuthally degenerate (e.g., in the black ellipses in the left panel). These $\ell$-$v$ bins are more likely to have a high metallicity scatter, as they probe the variations along different azimuths. For a galaxy with azimuthal metallicity variations, we expect that the $\ell$-$v$ bins with high absolute LSR velocities will have lower metallicity scatter compared to bins at lower absolute LSR velocities. This is exactly what we see for the C\&D gas and YSPs in the Romeo simulation (Figure \ref{fig:Romeo_OH_scatter}). Once again, the metallicity trend is insensitive to the value of the required minimum number of particles in each bin for both C\&D gas and YSPs.

Gas is not rotating at a constant circular speed, however, in neither the MW nor the simulated FIRE-2 galaxies. Dynamical interactions with spiral arms or a central bar can cause non-circular motions in the gas of a galaxy, termed ``streaming motions'' \citep{Erroz-Ferrer_2015, Kim_2024}. The motions dominate H\,{\sc i} line profiles and become apparent when gas appears at locations in $\ell$-$v$ diagrams forbidden by circular motions alone \citep{Shane_Bieger-Smith_1966, Fujimoto_1968, Burton_1971, Burton_1978}. Rotational speeds less than or greater than the constant circular speed can mimic streaming motions in $\ell$-$v$ space. We define any motions that do not obey a flat rotation curve as ``anomalous'' motions and describe how we identify anomalous motions in $\ell$-$v$ space below.

In an idealized galaxy where all of the gas is moving at a single circular speed, a well-defined envelope in $\ell$-$v$ space known as the terminal velocity curve is produced (see red curve in Figure \ref{fig:MW_LV}). Gas moving faster than the terminal velocity, i.e., $v_{\rm LSR} > v_{\rm LSR,terminal}$ in the first quadrant ($0\degree < \ell < 90\degree$) or $v_{\rm LSR} < v_{\rm LSR,terminal}$ in the fourth quadrant ($-90\degree < \ell < 0\degree$), cannot be explained by a single circular speed and signals the presence of anomalous motions. Similarly, in the second ($90\degree < \ell < 180\degree$) and third ($-180\degree < \ell < -90\degree$) quadrants, where purely circular motion predicts only negative and positive LSR velocities, respectively, detecting gas with the opposite sign of velocity indicates anomalous motions. Note that this method cannot capture all anomalous motions in a galaxy because radial and tangential velocities cannot be calculated with only longitude and LSR velocity information. Nevertheless, this method still allows us to identify the subset of anomalous motions that are near the locus of tangent points in a galaxy, which still covers $R_{\rm gal}$ up to the solar orbit and azimuths from around $-$90$\degree$ to 90$\degree$. For gas with anomalous motions in the first and fourth quadrants, we define the absolute difference between their LSR velocity and the terminal velocity as their ``excess $v_{\rm LSR}$'', or $v_{\rm LSR,excess}$. For gas with anomalous motions in the second and third quadrants, their excess $v_{\rm LSR}$ is defined as the absolute difference between their LSR velocity and zero.

\begin{figure}
    \centering
    \includegraphics[width=1\linewidth]{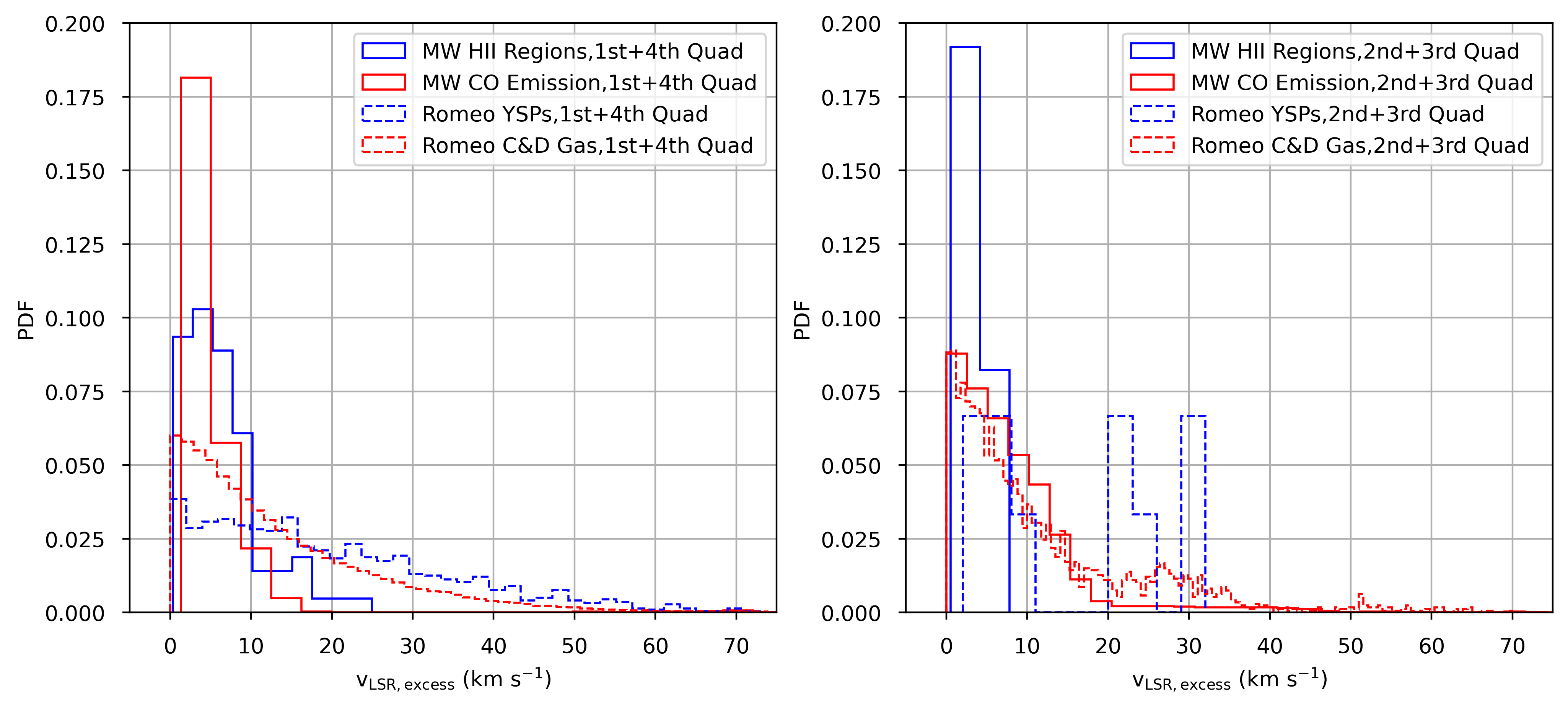}
    \caption{PDFs of $v_{\rm LSR,excess}$ for H\,{\sc ii} regions and CO emission in the MW and C\&D gas cells and YSPs in the Romeo simulation. Left panel displays the PDFs for the 1st and 4th galactic quadrants, and right panel displays the PDFs for the 2nd and 3rd galactic quadrants. See text for definition of $v_{\rm LSR,excess}$. In all four quadrants of Romeo, there are a few YSPs and C\&D gas cells with $v_{\rm LSR,excess}$ beyond 75 km s$^{-1}$ not shown here. Only CO emission with Galactic latitude $|b|\leq3\degree$ and integrated brightness temperatures, ${\rm log}_{10}(T)$, higher than the noise threshold, 0.5 K, is included in this plot. The mock observer for in the Romeo simulated is located at $(R_{\rm obs},\theta_{\rm obs})=(12.71$ kpc, $90\degree)$. H\,{\sc ii} regions, CO emission, C\&D gas, and YSPs that do not exhibit anomalous motions as defined in the text above are excluded from this plot.}
    \label{fig:excess_vlsr_hist}
\end{figure}

In the MW, 4.8\% of the H\,{\sc ii} regions and 12.77\% of the CO emission exhibit anomalous motions (Figure \ref{fig:MW_LV}), whereas in the Romeo simulation for a mock observer placed at $(R_{\rm obs},\theta_{\rm obs})=(12.71$ kpc, $90\degree)$, this percentage is 10.5\% for the YSPs and 6.7\% for the C\&D gas (Figure \ref{fig:Romeo_OH_abundances}). CO emission with an integrated brightness temperature ${\rm log}_{10}(T)$ less than the noise threshold of 0.5 K or with Galactic latitude $|b|>3\degree$ is excluded from this calculation. Figure \ref{fig:excess_vlsr_hist} shows the probability density functions (PDFs) of $v_{\rm LSR,excess}$ for the MW H\,{\sc ii} regions, MW CO emission, Romeo YSPs, and Romeo C\&D gas cells in the 1st and 4th galactic quadrants (left panel) and in the 2nd and 3rd galactic quadrants (right panel). We show PDFs instead of histograms since there are significantly fewer MW H\,{\sc ii} regions and less CO emission than YSPs and C\&D gas cells in Romeo. The bin sizes for the MW H\,{\sc ii} regions and Romeo YSPs in the 2nd and 3rd galactic quadrants are larger than those in the 1st and 4th galactic quadrants because the 2nd and 3rd galactic quadrants contain significantly fewer points in these two populations than the 1st and 4th galactic quadrants. The Romeo simulation has significantly larger anomalous motions than the MW: in both quadrant combinations, the Romeo YSPs and C\&D gas have longer $v_{\rm LSR,excess}$ tails compared to the MW H\,{\sc ii} regions and CO emission, respectively. In the 1st and 4th galactic quadrants, the Romeo YSP and C\&D gas velocity tails extend up to $v_{\rm LSR,excess} \simeq$ 60 km s$^{-1}$ and $v_{\rm LSR,excess} \simeq$ 50 km s$^{-1}$, respectively, compared to the MW H\,{\sc ii} region and CO emission velocity tails extending up to $v_{\rm LSR,excess}\simeq$ 25 km s$^{-1}$ and $v_{\rm LSR,excess}\simeq$ 15 km s$^{-1}$, respectively. In the 2nd and 3rd galactic quadrants, the high ends of the Romeo YSP and C\&D gas velocity distributions are at $v_{\rm LSR,excess} \simeq$ 30 km s$^{-1}$ and $v_{\rm LSR,excess} \simeq$ 35 km s$^{-1}$, respectively, compared to the high ends for the MW H\,{\sc ii} region and CO emission velocity distributions at $v_{\rm LSR,excess}\simeq$ 8 km s$^{-1}$ and $v_{\rm LSR,excess}\simeq$ 20 km s$^{-1}$, respectively. When comparing the 1st and 4th quadrants to the 2nd and 3rd quadrants, the Romeo YSPs and C\&D gas have longer velocity tails in the 1st and 4th quadrants than in the 2nd and 3rd quadrants, whereas the MW H\,{\sc ii} region and CO emission velocity distributions in the two quadrant combinations are similar. Based on these comparisons, the kinematic discrepancies between Romeo simulation and the MW are strongest in the 1st and 4th quadrants. Varying $R_{\rm obs}$ and $\theta_{\rm obs}$ of the mock observer in the Romeo simulation and using other MW-mass galaxies from the FIRE-2 simulations yields similar results.

\begin{figure}
    \centering
    \includegraphics[width=1\linewidth]{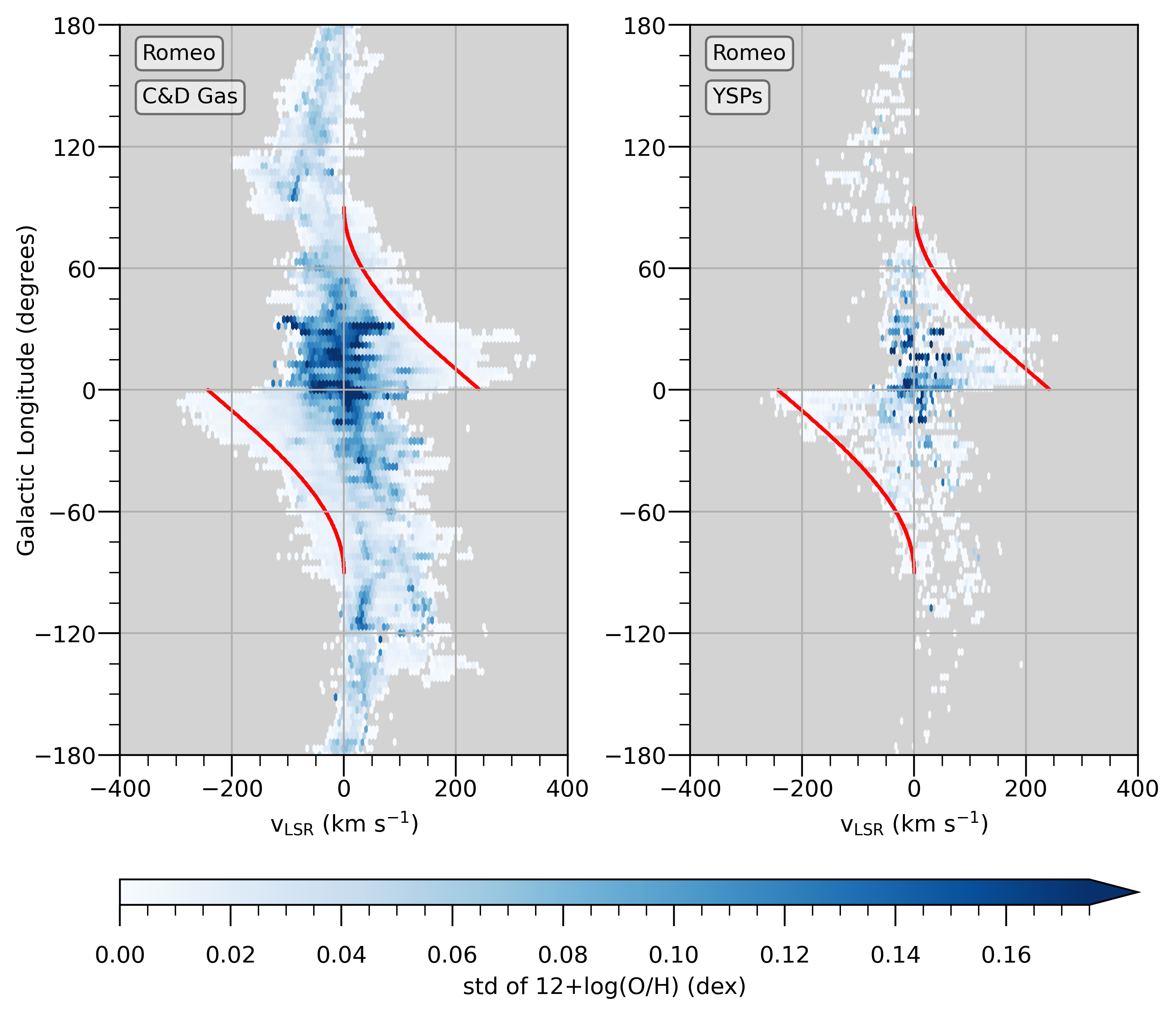}
    \caption{Oxygen abundance scatter in the Romeo simulation at z=0 after artificially removing azimuthal variations. Figure details are similar to Figure \ref{fig:Romeo_OH_scatter}. The C\&D gas is in the left panel and YSPs in the right panel. Note that even with no azimuthal variations, there is a higher oxygen abundance scatter for bins near $v_{\rm LSR}=0$ compared to bins near the terminal velocity in both panels. Large anomalous motions present in the Romeo simulation reproduce this metallicity trend and mimic azimuthal variations in $\ell$-$v$ space (see text).
}
    \label{fig:Romeo_OH_scatter_no_azi_vars}
\end{figure}

These large anomalous motions mimic azimuthal structure in $\ell$-$v$ space. Figure \ref{fig:Romeo_OH_scatter_no_azi_vars} shows the $\ell$-$v$ diagram of the Romeo simulation after removing all azimuthal metallicity variations. These variations were removed by fitting a linear radial metallicity gradient to the C\&D gas and overriding the metallicity of the gas with the metallicity of the fitted gradient. The exact details of this fit are unimportant to Figure \ref{fig:Romeo_OH_scatter_no_azi_vars} as long as the metallicity of the C\&D gas obeys a perfect radial gradient. Despite the removal of the azimuthal variations, the metallicity trend as described above still persists: $\ell$-$v$ bins with high $|v_{\rm LSR}|$ have a lower oxygen abundance scatter than bins with low $|v_{\rm LSR}|$. The reason is that the large anomalous motions in the Romeo simulation cause $\ell$-$v$ bins with low $|v_{\rm LSR}|$ to actually probe gas at widely different $R_{\rm gal}$. The oxygen abundance scatter will still be high as the standard deviation is calculating the scatter of radial metallicity variations. Therefore, if a galaxy contains large anomalous motions, it is impossible to tell if the metallicity trend in $\ell$-$v$ space is due to azimuthal variations or anomalous motions.

\begin{figure}
    \centering
    \includegraphics[width=1\linewidth]{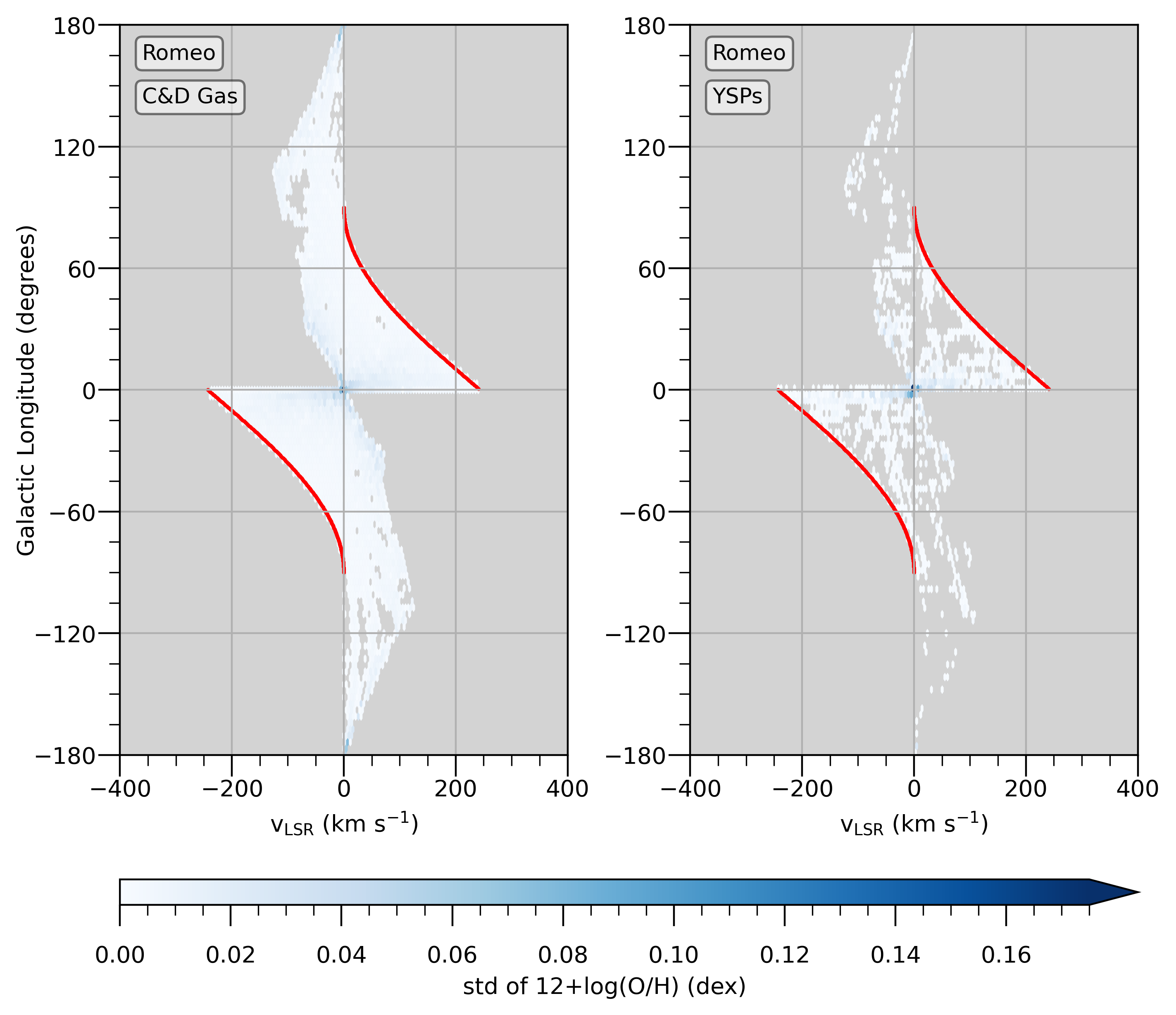}
    \caption{Oxygen abundance scatter in the Romeo simulation after artificially removing all azimuthal variations and anomalous motions. Figure details are similar to Figure \ref{fig:Romeo_OH_scatter}. The C\&D gas is in the left panel and YSPs in the right panel. With neither azimuthal variations nor anomalous motions, the oxygen abundance scatter is near zero for all $\ell$-$v$ bins.}
    \label{fig:Romeo_OH_scatter_no_azi_vars_no_anom_motions}
\end{figure}

To remove the oxygen abundance scatter for all $\ell$-$v$ bins, both anomalous motions and azimuthal variations in a galaxy need to be removed. The absence of anomalous motions ensures that bins with low $|v_{\rm LSR}|$ probe gas at similar $R_{\rm gal}$ and the absence of azimuthal variations ensures that the metallicity of gas at the same $R_{\rm gal}$ is constant. Figure \ref{fig:Romeo_OH_scatter_no_azi_vars_no_anom_motions} shows the $\ell$-$v$ diagram of the oxygen abundance scatter after both azimuthal variations and anomalous motions are removed for C\&D gas and YSPs in the Romeo simulation. As expected, all $\ell$-$v$ bins have an oxygen abundance scatter near zero. Maintaining anomalous motions but keeping them to a minimum also reproduces the near zero oxygen abundance scatter in $\ell$-$v$ space (not shown).

\begin{figure}
    \centering
    \includegraphics[width=1\linewidth]{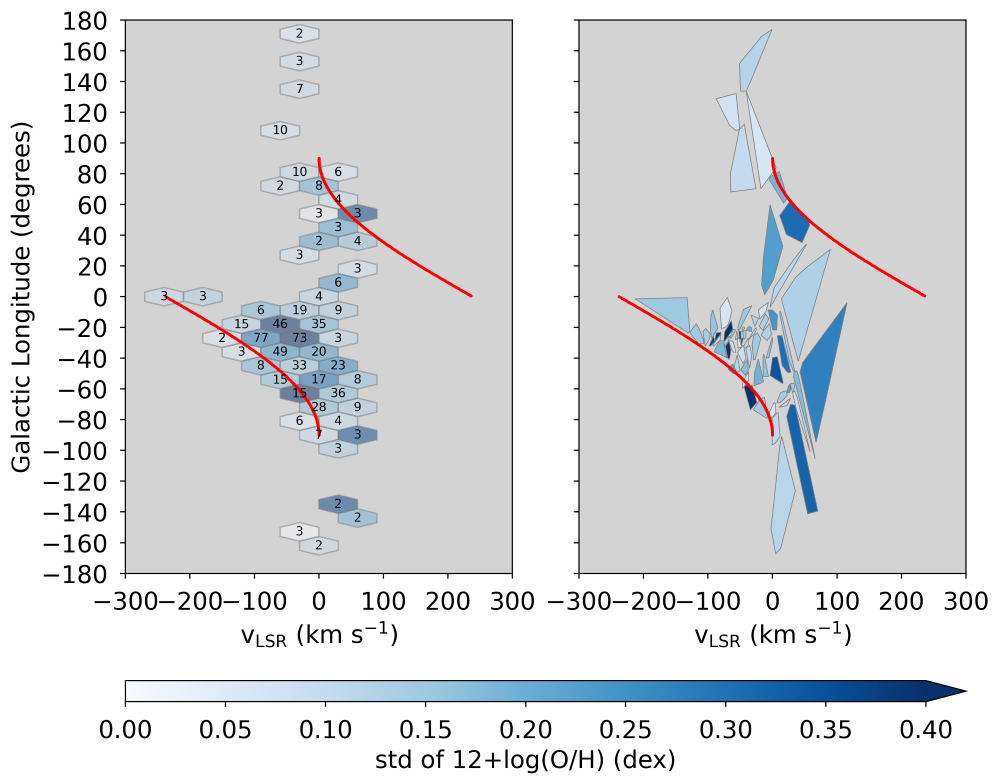}
    \caption{Azimuthal metallicity variations in the MW using H\,{\sc ii} regions from the WISE Catalog. The two panels display $\ell$-$v$ diagrams with different binning schemes (see caption of Figure \ref{fig:MW_abundances} for details). Both panels plot the standard deviation of the oxygen abundance in each bin.}
    \label{fig:MW_scatter}
\end{figure}

The MW has relatively small anomalous motions compared to the simulated Romeo galaxy (see Figure \ref{fig:excess_vlsr_hist}). This is supported by studies of H\,{\sc i} gas, CO gas, and stars in the MW, which have streaming motions up to $\sim$10 km s$^{-1}$ \citep{McClure-Griffiths_2007, Khanna_2023, Soler_2025}. Figure \ref{fig:MW_scatter} shows the binned $\ell$-$v$ diagram of known MW H\,{\sc ii} regions colored by the oxygen abundance scatter within each bin. Azimuthal variations may be present, as bins with $-60\degree \leq \ell \leq -25\degree$ exhibit increased oxygen abundance scatter moving from negative $v_{\rm LSR}$ to zero $v_{\rm LSR}$. In the Southern hemisphere, the effects of velocity crowding dominate and complicate interpretation outside this longitude range ($-90 \leq \ell \leq -75\degree$ and $-10 \leq \ell \leq 10\degree$), while in the Northern hemisphere, the number of H\,{\sc ii} regions is insufficient. The metallicity trend at $-60\degree \leq \ell \leq -25\degree$ is smaller than in the Romeo simulation (see Figure \ref{fig:Romeo_OH_scatter}). In the Milky Way, however, disentangling the influence of azimuthal metallicity fluctuations from that of anomalous gas motions on the metallicity trend is difficult, as the magnitude of anomalous gas motions in the Milky Way are not well constrained. In theory, we can artificially remove all azimuthal variations in the MW by fitting a linear function to the oxygen abundances and $R_{\rm gal}$ and then overriding the oxygen abundance of all H\,{\sc ii} regions by that of the fit. This procedure, however, neglects that $R_{\rm gal}$ is derived assuming strictly circular motions for gas, and thus implicitly assumes there are no anomalous gas motions in the MW. 

\begin{figure}
    \centering
    \includegraphics[width=1\linewidth]{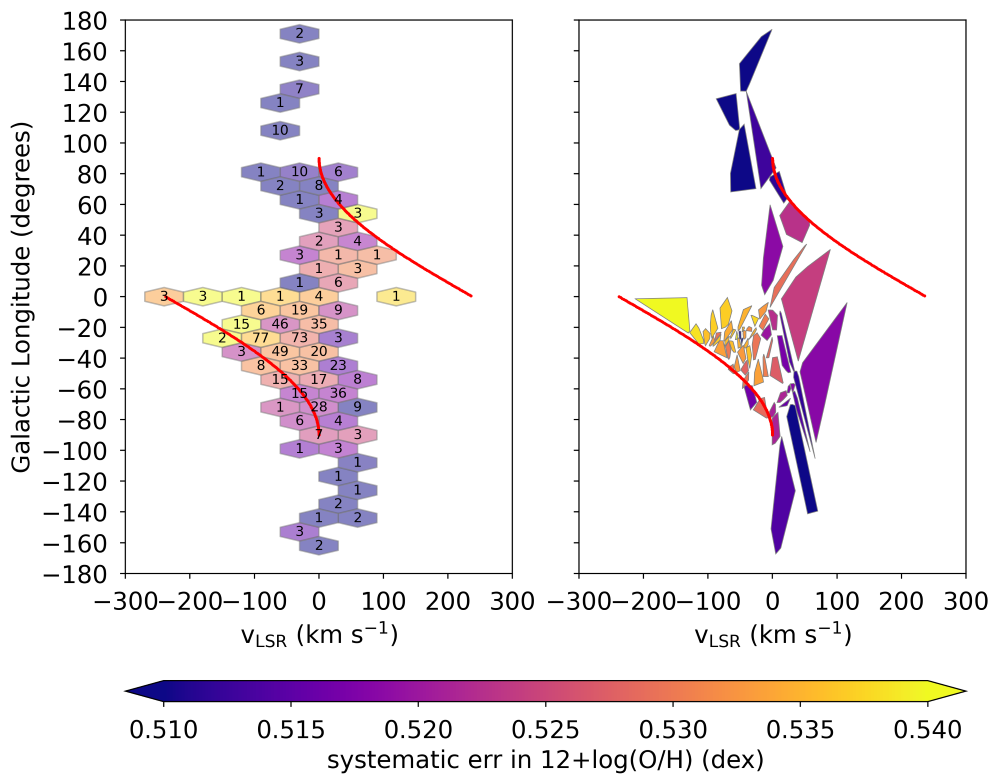}
    \caption{Mean systematic metallicity uncertainty in the MW using H\,{\sc ii} regions from the WISE Catalog. The two panels display $\ell$-$v$ diagrams with different binning schemes (see caption of Figure \ref{fig:MW_abundances} for details). Both panels plot the mean systematic error of the oxygen abundance in each bin. The systematic errors in the oxygen abundance are still visibly higher than the scatters in the oxygen abundance, which casts doubt on the detection of azimuthal variations in Figure \ref{fig:MW_scatter}.}
    \label{fig:MW_err}
\end{figure}

Furthermore, the scatter in each bin is smaller than the error, with the scatter typically between 0.00 to 0.40 dex and the error between 0.51 to 0.54 dex. Figure \ref{fig:MW_err} shows the mean systematic oxygen abundance error within each bin for the MW. The systematic errors are shown instead of the total errors because 82 sources lack derived observational electron temperature errors, preventing a calculation of the total oxygen abundance errors for these sources. Regardless, this is not an issue since the systematic errors dominate over the observed errors. The systematic oxygen abundance errors are visibly larger than the scatter, so the scatter in Figure \ref{fig:MW_scatter} may be entirely explained by these errors. As a reminder, these systematic errors can be in principle largely reduced by applying correction factors to the oxygen abundances of these H\,{\sc ii} regions.

\section{Discussion} \label{sec:discussion}
Using $\ell$-$v$ diagrams, we can detect the presence of radial metallicity gradients and characterize whether they are positive, flat, or negative with increasing $R_{\rm gal}$, but not the magnitude of the gradient's slope. As such, the $\ell$-$v$ diagram can only qualitatively describe the radial gradient. Regardless, even the sign of the radial gradient is informative about a galaxy's growth, star formation history, and stellar feedback. A positive gradient could indicate a history of galactic fountaining, which is when metal-rich gas is expelled from the galactic center due to strong outflows and then re-accreted onto the disk's outskirts (e.g., \citealt{Schonrich_2017}). A flat gradient indicates strong mixing, probing processes such as central gas outflows or radial migration of stars that could dilute the central metallicity (e.g., \citealt{Sharda_2021}). A negative gradient indicates higher star formation rates in the inner disk than the outer disk, probing star formation (e.g., \citealt{Graf_2024}).

Do we see non-axisymmetric azimuthal variations in the MW? \citet{Orr_etal_2023_metalfreeways} proposed that large-scale non-circular gas motions driven by Galactic spiral arms and/or bars create gas-phase azimuthal variations. \citet{Grand_etal_2016} suggests that stellar radial migration induced by spiral arm perturbations is the cause for azimuthal variations in specifically young stars, and  \citet{Khoperskov_2018} offers that the trapping of young, metal-rich stars in the gravitational potential of spiral arms is the cause for azimuthal variations across all stars. In the MW, the detection of such variations in $\ell$-$v$ space is uncertain, largely because the oxygen abundance systematic uncertainties exceed the expected scatter signal. These systematic effects can be reduced by applying correction factors to the oxygen abundances. Doing so, however, requires deriving the electron density of the nebular region and spectral type of the ionizing star from the RRLs, radio continuum, and distance to the H\,{\sc ii} region \citep{Balser_2024}. This calculation is beyond the scope of this paper. Additionally, some regions in $\ell$-$v$ space are sparsely populated, such as in the Northern hemisphere ($\ell > 0\degree$), making it difficult to detect a trend in the oxygen abundance scatter of $\ell$-$v$ bins. Obtaining more sources for these sparsely populated regions would increase the number of bins in these regions and improve detectability of azimuthal variations.

We have also found that the FIRE-2 simulated MW-mass galaxies have significantly different kinematics from the MW. These simulated galaxies have significantly larger anomalous motions, defined as radial streaming motions and circular velocities deviant from a constant rotation speed, than those in the MW (see Figure \ref{fig:excess_vlsr_hist}). Specifically, the excess LSR velocity tails for C\&D gas and YSPs in the simulated FIRE-2 galaxies reach up to $\sim$60 km s$^{-1}$, whereas the tails only go up to $\sim$25 km s$^{-1}$ for H\,{\sc ii} regions and CO emission in the MW. Additionally, \citet{McCluskey_2023} show that the stellar populations in the FIRE-2 simulations are dynamically hotter than those in the MW. These velocity discrepancies could be due to certain physical processes not being modeled in the FIRE-2 simulations. For example, the simulations do not include AGN feedback, magnetohydrodynamics, anisotropic conduction and viscosity, self-consistent cosmic-ray injection and transport, self-consistent creation and destruction of dust, nor radiation hydrodynamics via methods such as flux-limited diffusion or the M1 radiative transfer model. With regards to galaxy-wide properties, the lack of these physics has only a weak effect or no effect at all \citep{Wetzel_2023_FIRE2DataRelease}. It is possible, however, that these missing physics subtly affect the observables of sub-populations in these simulations, such as the kinematics of C\&D gas and YSPs with anomalous motions. The specific treatment of stellar feedback may also play a role in why the simulated galaxies exhibit larger anomalous motions compared to the MW \citep{Hopkins_etal_2018_FIRE, Wetzel_2023_FIRE2DataRelease}.

\begin{figure}
    \centering
    \includegraphics[width=1\linewidth]{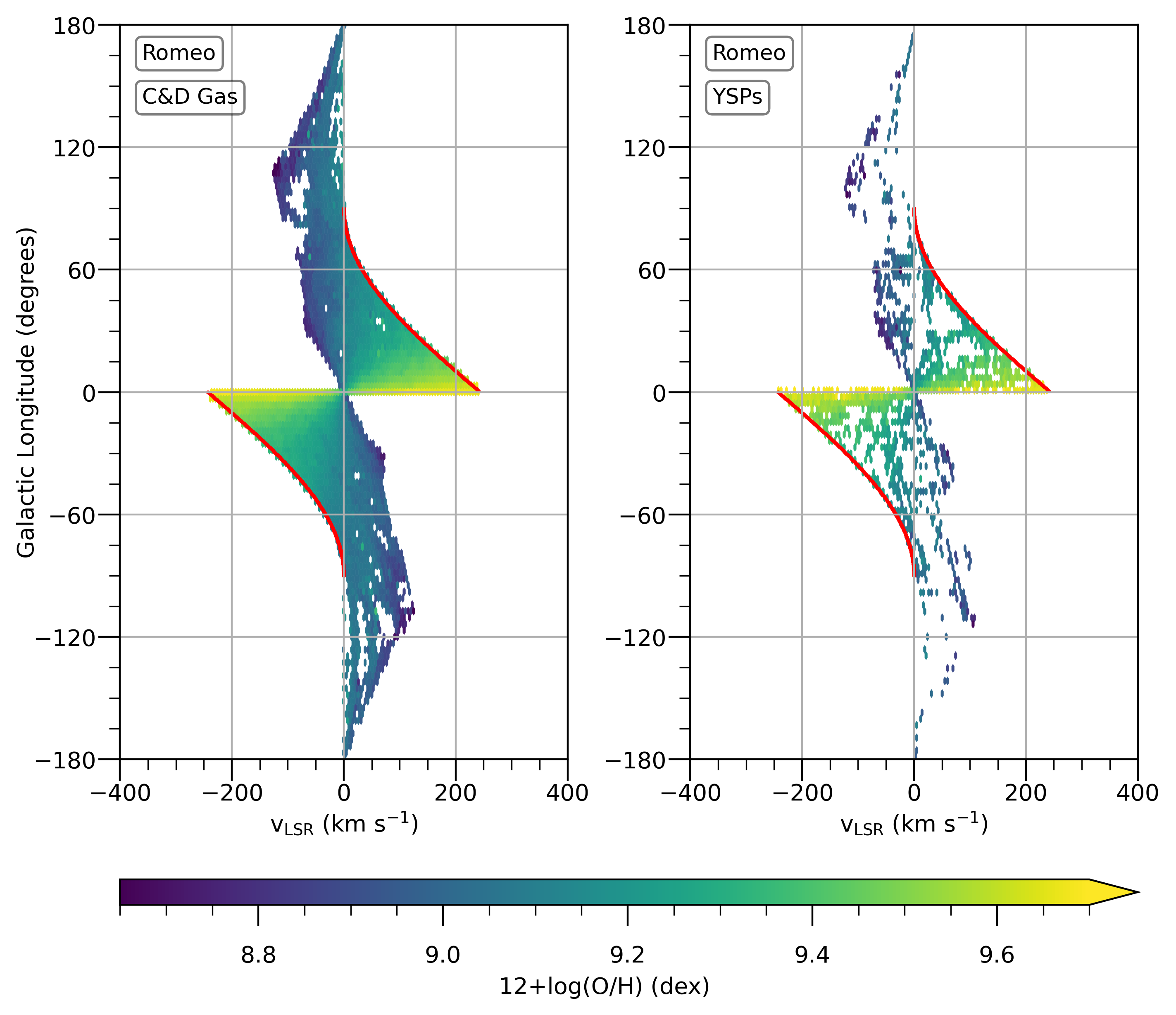}
    \caption{Oxygen abundances in the Romeo simulation at z=0 after artificially removing anomalous motions. Figure details are similar to Figure \ref{fig:Romeo_OH_scatter}. The C\&D gas is in the left panel and YSPs in the right panel. The terminal velocity curve method and constant galactic longitude method both still detect the radial metallicity gradient for both the C\&D gas and the YSPs.}
    \label{fig:Romeo_OH_abundances_no_anom_motions}
\end{figure}

Do these anomalous motions affect the detection of metallicity structure in $\ell$-$v$ space? Our interpretation of the radial metallicity gradient in $\ell$-$v$ space assumes that Galactic kinematics obey GRMs with no radial motions, but we know this is not true in the MW and the FIRE-2 simulations \citep{Erroz-Ferrer_2015, Kim_2024}. In Figure \ref{fig:Romeo_OH_abundances_no_anom_motions}, we explore the effect of removing the large anomalous motions in the Romeo simulation on the $\ell$-$v$ diagram of oxygen abundance. This plot shows that anomalous motions cannot mimic the radial gradient, as both the terminal velocity curve method and the constant galactic longitude method still detect the radial gradient. Large anomalous motions do, however, mimic azimuthal structure in $\ell$-$v$ space and cause a false detection (see Figures \ref{fig:Romeo_OH_scatter}, \ref{fig:Romeo_OH_scatter_no_azi_vars}, and \ref{fig:Romeo_OH_scatter_no_azi_vars_no_anom_motions}).

One potential concern about our analysis is that these metallicity trends might appear only in the Romeo galaxy. To address this, we examine other MW-mass galaxies at z=0 from the Latte and ELVIS suites. These galaxies also display the negative gradient and azimuthal variations in their $\ell$-$v$ diagrams (see Figures \ref{fig:lv_multiple_galaxies} and \ref{fig:lv_std_multiple_galaxies} in Appendix \ref{sec:appendix_lv_diagrams}).

Another concern is that $\ell$-$v$ diagrams are sensitive to the observer's position within the galaxy. To test this, we vary the observer's galactic azimuth and galactocentric radius in the Romeo galaxy and generate the corresponding $\ell$-$v$ diagrams (see Figures \ref{fig:lv_multiple_angles}, \ref{fig:lv_std_multiple_angles}, \ref{fig:lv_multiple_rgal}, and \ref{fig:lv_std_multiple_rgal} in Appendix \ref{sec:appendix_lv_diagrams}). These diagrams consistently show the negative radial metallicity gradient and azimuthal variations at all observer locations, demonstrating that these trends are not artifacts of a specific vantage point. 

We also explore other metallicity tracers that have been used when studying the chemical structure of the MW. Both nitrogen (N) and iron (Fe) have been used as oxygen (O) alternatives when studying the MW and testing Galactic chemical evolution models (e.g., N/H, \citealt{Pineda_2024, Zhang_2025}; Fe/H, \citealt{Grand_etal_2016, Bellardini_2022, Graf_2024}; Fe/H and N/H, \citealt{Bellardini_etal_2021}). \citet{Bellardini_etal_2021} showed that FIRE-2 MW-mass galaxies have N/H and Fe/H (and O/H) radial gradients and azimuthal variations similar to those observed in nearby galaxies of similar mass. To compare to these studies, we test if the radial gradient and azimuthal variations are seen in $\ell$-$v$ space when using Fe or N as the metallicity tracer instead of O in the FIRE-2 simulations. We find that the $\ell$-$v$ diagrams of metallicity and metallicity variations with Fe and N (not shown) are almost identical to those with O. 

Additionally, we test if weighting the metallicity by mass of the star particles or gas cells when binning the O/H abundance ratio has an effect on detecting metallicity structure in $\ell$-$v$ space. Figure \ref{fig:mass_dist} shows the distribution of masses for the C\&D gas and the YSPs in Romeo at z=0. In the C\&D gas distribution, $\sim$96.72\% have masses between 3500 $M_{\rm sun}$ and 10000 $M_{\rm sun}$, while the remaining $\sim$3.28\% have masses between 10000 $M_{\rm sun}$ and 145000 $M_{\rm sun}$. For the YSPs, $\sim$94.47\% have masses between 2500 $M_{\rm sun}$ and 10000 $M_{\rm sun}$, and the remaining $\sim$5.53\% have masses between 10000 $M_{\rm sun}$ and 40000 $M_{\rm sun}$. These high-mass outliers could potentially skew the mean and the scatter of the O/H abundance ratio, leading to artificial detections of the radial gradient and azimuthal variations in $\ell$-$v$ space. In actuality, however, the $\ell$-$v$ diagrams of the mass-weighted mean O/H abundance ratio and the mass-weighted scatter of the O/H abundance ratio look nearly identical to their non-mass-weighted counterparts, thus showing that the detection of metallicity structure in $\ell$-$v$ space in the FIRE-2 simulations is robust to outliers with extreme masses. The same analysis performed on the other five simulated FIRE-2 galaxies in this paper yield the same result.

\citet{Wetzel_2023_FIRE2DataRelease} note that the FIRE-2 simulations tend to moderately overestimate $\alpha$-element abundances (like O, Ca, and Mg). The lack of active galactic nuclei (AGNs) in the Latte and ELVIS simulations may cause the higher FIRE-2 oxygen abundances in the C\&D gas and YSPs, as these simulations lack star formation quenching from AGN outflows. The star formation rate across these simulations' histories may be more elevated compared to the MW's history, resulting in an enriched ISM. The nucleosynthetic yield tables also impact the oxygen abundances. The FIRE-2 simulations use the yields in \citet{Nomoto_2006} for core-collapse supernovae, \citet{Iwamoto_1999} for Type Ia supernovae, and the models from \citet{van_den_Hoek_1997}, \citet{Marigo_2001}, and \citet{Izzard_2004} as synthesized in \citet{Wiersma_2009} for OB/AGB winds. All these yields are averaged over the initial mass function from \citet{Kroupa_2001}. The O/H abundance ratios in the FIRE-2 simulations are systematically $\sim$0.5 dex higher than in MW H\,{\sc ii} regions. Nevertheless, the absolute oxygen abundance does not matter for our analysis but rather the radial metallicity gradient, which is still detected in both the MW's and the FIRE-2 galaxies' $\ell$-$v$ diagrams.

The FIRE-2 simulations are a great tool for understanding how azimuthal metallicity variations formed in the MW. From analyzing the gas-phase kinematics in six simulated FIRE-2 MW-mass galaxies, \citet{Orr_etal_2023_metalfreeways} offer that the origin of azimuthal metallicity variations is due to the movement of gas along spiral arms as the arms move radially inward and outward, causing the metallicity of these arms to mix with the surrounding ISM (also see \citealt{Graf_2025}). If the MW's radial motions are smaller than those in the FIRE-2 simulations, however, then the proposed mechanism for producing azimuthal metallicity structure will be less effective. This may be possible, as the MW contains significantly smaller anomalous motions compared to the FIRE-2 simulations (see Figure \ref{fig:excess_vlsr_hist}). There may be multiple processes at play that combine to create the observed azimuthal metallicity variations. Spiral arms could induce stellar radial migration \citep{Grand_etal_2016} and trap young, metal-rich stars \citep{Khoperskov_2018} to help create azimuthal metallicity variations. Other tracers of chemical evolution are required to distinguish between these models.

\section{Summary}
Measuring the chemical structure of galactic disks is essential to understanding the host galaxy's formation and evolution. The radial metallicity gradient and non-axisymmetric azimuthal variations in both the MW and other galaxies have been used extensively to constrain models of star formation history, stellar migration, stellar feedback, and radial gas inflows. H\,{\sc ii} regions are powerful tracers of chemical structure in the MW. RRLs from H\,{\sc ii} regions are among the few tracers unaffected by dust, allowing them to probe the full extent of the Galactic disk. Since metals act as coolants, they primarily regulate the thermal motions of ionized gas, resulting in a linear correlation between metallicity and electron temperature. Assuming LTE, the ratio of the RRL to radio free-free continuum offers a measure of the electron temperature independent of electron density, thus providing an indirect method to estimate metallicity \citep{Wenger_2019_VLA_sample}.

Most studies of the radial metallicity gradient and azimuthal variations in the MW construct the spatial distribution of metallicities from distance estimates to sources. Kinematic distance determinations, however, are subject to the kinematic distance ambiguity and velocity crowding. While parallax-determined distances to H\,{\sc ii} regions provides some constraints to kinematic distances (e.g.. \citealt{Wenger_2018}), to avoid these challenges, we analyze the metallicity structure of the MW using only $\ell$-$v$ diagrams. We further extend this analysis to MW-mass galaxies from the FIRE-2 simulations, where we find that $\ell$-$v$ diagrams recover both the negative radial metallicity gradient and azimuthal variations in C\&D gas and YSPs, independent of observer position and metallicity tracer. In the MW, the $\ell$-$v$ diagram constructed from WISE H\,{\sc ii} region data clearly shows the negative radial gradient, but the high oxygen abundance errors prevent any detection of azimuthal variations. These errors can be largely reduced by applying correction factors to the oxygen abundances, but this procedure requires knowledge of the distances, which we avoid invoking in this paper.

Additionally, we find that the MW-mass galaxies from the FIRE-2 simulations exhibit significantly different gas kinematics compared to the MW, with much larger anomalous motions than seen in the Milky Way. Prior studies have also noted discrepancies between the velocity dispersions of the FIRE-2 simulated galaxies versus the MW. These discrepancies may stem from the absence of AGN feedback in the FIRE-2 simulations and necessitate caution when interpreting drivers of azimuthal metallicity variations using the kinematics in these simulations. Anomalous motions cannot mimic radial metallicity gradients in $\ell$-$v$ space. If the motions are large enough, however, they do produce artificial azimuthal variations in $\ell$-$v$ space, as seen in the Romeo simulation.

We acknowledge that $\ell$-$v$ diagrams can only provide qualitative results about metallicity structure in galaxies, whereas previous studies that use distances provide quantitative results. \textit{Despite this, $\ell$-$v$ diagrams serve as a novel complementary tool to study metallicity structure and can even detect azimuthal metallicity variations if Galactic anomalous motions are small enough. To our knowledge, this is the first time $\ell$-$v$ diagrams have been used to probe metallicity structure.}

\begin{acknowledgements}
We are thankful to Andrew Wetzel and Butler Burton, whose comments and feedback greatly improved the manuscript. We also thank Matthew Orr, Elizabeth Iles, and Joshua Peek for helpful discussions about the FIRE-2 simulations. We are grateful to Cosima Eibensteiner for providing valuable discussions and references on typical radial velocities in MW-like galaxies. T.V.W. is supported by a National Science Foundation Astronomy and Astrophysics Postdoctoral Fellowship under award AST-2202340. The National Radio Astronomy Observatory is a facility of the National Science Foundation operated under cooperative agreement by Associated Universities, Inc.

We use H\,{\sc ii} region data from \citep{Wenger_2021_WISE_catalog}, which is a SQL database of ionized gas data (mostly radio recombination lines) toward nebulae in the WISE Catalog of Galactic H\,{\sc ii} Regions \citep{Anderson_2014}. We also use simulations from the FIRE-2 public data release \citep{Wetzel_2023_FIRE2DataRelease}. The FIRE-2 cosmological zoom-in simulations of galaxy formation are part of the Feedback In Realistic Environments (FIRE) project, generated using the Gizmo code \citep{Hopkins_2015_GIZMO} and the FIRE-2 physics model \citet{Hopkins_etal_2018_FIRE}.

\software{Matplotlib \citep{matplotlib_2007}, pandas \citep{pandas_2010}, NumPy \& SciPy \citep{numpy_scipy_2011}, Astropy \citep{astropy_2013}, gizmo\_analysis \citep{GizmoAnalysis_2020} (first used in \citealt{Wetzel_2016_m12i}), sqlite3 \citep{sqlite_2020}}
\end{acknowledgements}

\appendix
\section{Production of $\ell$-$v$ Diagrams in FIRE-2 Simulations} \label{sec:appendix_equations}
FIRE-2 uses a particle-based method where stars are represented by discrete particles and gas is represented by volumetric cells, allowing for detailed tracking of their individual motions. The models incorporate various physics like the heating and cooling of gas in the multiphase interstellar medium, stellar feedback (from supernovae and stellar winds), and radiative feedback from hot gas, which all significantly impact the kinematics of the gas and stars. By integrating the forces acting on each particle/cell over time, the simulation directly calculates their velocities at each time step. 

In order to produce the $\ell$-$v$ diagrams from simulations, we need to calculate the galactic longitude and line-of-sight velocity at a given observer position from the positions and velocities of the particles/cells. In all of the following equations, we ignore the z-component of the positions and velocities and set the height of the observer at a height of $z=0$ so the galactic midplane is viewed at an inclination of 0 degrees.

To get the galactic longitudes $\ell$, we first transform the coordinate system from one centered on the galactic center to one centered on the observer position. We then rotate the coordinate system such that the positive x-axis is now the line connecting the observer to the galactic center. Finally, we calculate the longitudes $\ell$ in the new coordinate system.

The galactic longitude $\ell$ is given by
\begin{equation}
    \ell  = {\rm tan}^{-1} \biggl(\frac{q_{\rm p,y_{rot}}}{q_{\rm p,x_{rot}}} \biggr)* \biggl( \frac{180\degree}{\pi} \biggr)\,\,{\rm  degrees},
\end{equation}
where $\bar{q}_{\rm p}$ is the position of the particle in the observer coordinate system (defined as the Cartesian system where origin is placed at the observer's position and the positive x-axis is set to the line connecting the observer to the galactic center). The particle's projected distances along the x and y directions are $q_{\rm p,y_{rot}}$ and $q_{\rm p,x_{rot}}$, respectively. We use a rotation matrix to derive $\bar{q}_{\rm p}$ by
\begin{equation}
    \bar{q}_{\rm p} = R\times(\bar{p}_{\rm p}-\bar{p}_{\rm obs}),
\end{equation}
where $\bar{p}_{\rm p}$ and $\bar{p}_{\rm obs}$ are the position of the particle and the observer, respectively, in the galactic center coordinate system (defined as the Cartesian system where the origin is placed at the Galactic center and the x- and y-axes are set to the galaxy's principal axes). \textit{R} is the rotation matrix given by
\begin{equation}
    R = \begin{bmatrix}
    {\rm cos}\,(\theta_{\rm rot}) & -{\rm sin}\,(\theta_{\rm rot}) \\
    {\rm sin}\,(\theta_{\rm rot}) & {\rm cos}\,(\theta_{\rm rot})
    \end{bmatrix}.
\end{equation}
Here, $\theta_{\rm rot}$ is given by 
\begin{equation}
    \theta_{\rm rot} = {\rm tan}^{-1} \biggl (\frac{p_{\rm obs,x}}{p_{\rm obs,y}} \biggr).
\end{equation}
The observer position $\bar{p}_{\rm obs}$ for each FIRE-2 galaxy is determined by placing the observer at a radius akin to where the Sun is in the MW, but corrected for the virial radius in the FIRE-2 galaxy being different from that in the MW (see Table \ref{tab:sim_properties}). 

To get the LSR velocities, $v_{\rm LSR}$, we project the 2D velocity vector of the particle to the 2D unit vector $\hat{d}$ connecting the observer to the particle. We likewise project the observer's velocity (taken as purely circular motion) to $\hat{d}$, and then subtract out the projected gas velocity by the projected observer velocity to get $v_{\rm LSR}$. 

To calculate the LSR velocity, we use
\begin{equation}
    v_{\rm LSR} = (\bar{v}_{\rm p}\,\cdot\,\hat{d})-(\bar{v}_{\rm obs}\,\cdot\,\hat{d}),
\end{equation}
where $\bar{v}_{\rm p}$ and $\bar{v}_{\rm obs}$ are the velocity of the particle and of the observer in the galactic center coordinate system, respectively. The unit vector from the observer to the particle $\hat{d}$ is given by
\begin{equation}
    \hat{d} = \frac{\bar{p}_{\rm p} - \bar{p}_{\rm obs}}{||\bar{p}_{\rm p} - \bar{p}_{\rm obs}||},
\end{equation}
and $\bar{v}_{\rm obs}$ is derived from the observer's circular velocity by
\begin{equation}
    \bar{v}_{\rm obs} = -v_{\rm circ,obs}\,{\rm sin}\phi\,\hat{x}\,+v_{\rm circ,obs}\,{\rm cos}\phi\,\hat{y}\,\,,
\end{equation}
where $\phi$ is the angle from the positive x-axis to the observer in the galactic center coordinate system. We set the observer's circular velocity $v_{\rm circ,obs}$ to the mean circular velocity of the C\&D gas:
\begin{equation}
    v_{\rm circ,obs} = {\rm mean}(v_{\rm circ,C\&D})\,\,\,, \end{equation}
where we only consider C\&D gas located within $|x| \leq$ 20 kpc, $|y|\leq$ 20 kpc, and $|z|\leq$ 7.5 kpc. To graph the terminal velocity curves, we use Equation 4.2 from \citet{Burton_1974_thesis}:
\begin{equation} \label{eq:burton}
    v_{\rm LSR,terminal} = (\omega_{\rm p} - \omega_{\rm obs})\,R_{\rm obs}\,{\rm sin}(\ell)\,\,,
\end{equation}
where $\omega_{\rm p}$ and $\omega_{\rm obs}$ are the angular velocity of the particle and of the observer, respectively, and $R_{\rm obs}$ is the galactocentric radius of the observer. The angular velocity of the particle $\omega_{\rm p}$ is given by
\begin{equation}
    \omega_{\rm p} = \frac{v_{\rm circ,p}}{r_{\rm p}}\,\,,
\end{equation}
where $r_{\rm p}$ is the galactocentric radius of the particle given by
\begin{equation}
    r_{\rm p} = R_{\rm obs}|sin(\ell)|\,\,.
\end{equation}
Combining the above two equations, we get that
\begin{equation}
    \omega_{\rm p} =\frac{v_{\rm circ,p}}{R_{\rm obs}|{\rm sin}(\ell)|}\,\,.
\end{equation}
Note that $v_{\rm circ}$ is the circular velocity in km s$^{-1}$, while $\omega$ is the angular velocity in radians s$^{-1}$. The angular velocity of the observer $\omega_{\rm obs}$ is given by
\begin{equation}
    \omega_{\rm obs} = \frac{v_{\rm circ,obs}}{R_{\rm obs}}\,\,.
\end{equation}
We set the circular velocities of the particles and the observer to the mean circular velocity of the C\&D gas in that galaxy to get a constant rotation speed, which then is used for deriving the terminal velocities. Explicitly stated,
\begin{equation}
    v_{\rm circ,p}=v_{\rm circ,obs}={\rm mean}(v_{\rm circ,C\&D})\,\,.
\end{equation}
Substituting everything into Equation \ref{eq:burton}, we get
\begin{equation}
     v_{\rm LSR,terminal}= {\rm mean}(v_{\rm circ,C\&D})\,\,{\rm sin}(\ell)\,\biggl( \frac{1}{|{\rm sin}(\ell)|} - 1 \bigg)\,\,.
\end{equation}
This final expression for the terminal velocity depends on only two variables.

\section{$\ell$-$v$ Diagrams in FIRE-2 Simulations} \label{sec:appendix_lv_diagrams}
\begin{figure}[h]
    \centering
    \includegraphics[width=1\linewidth]{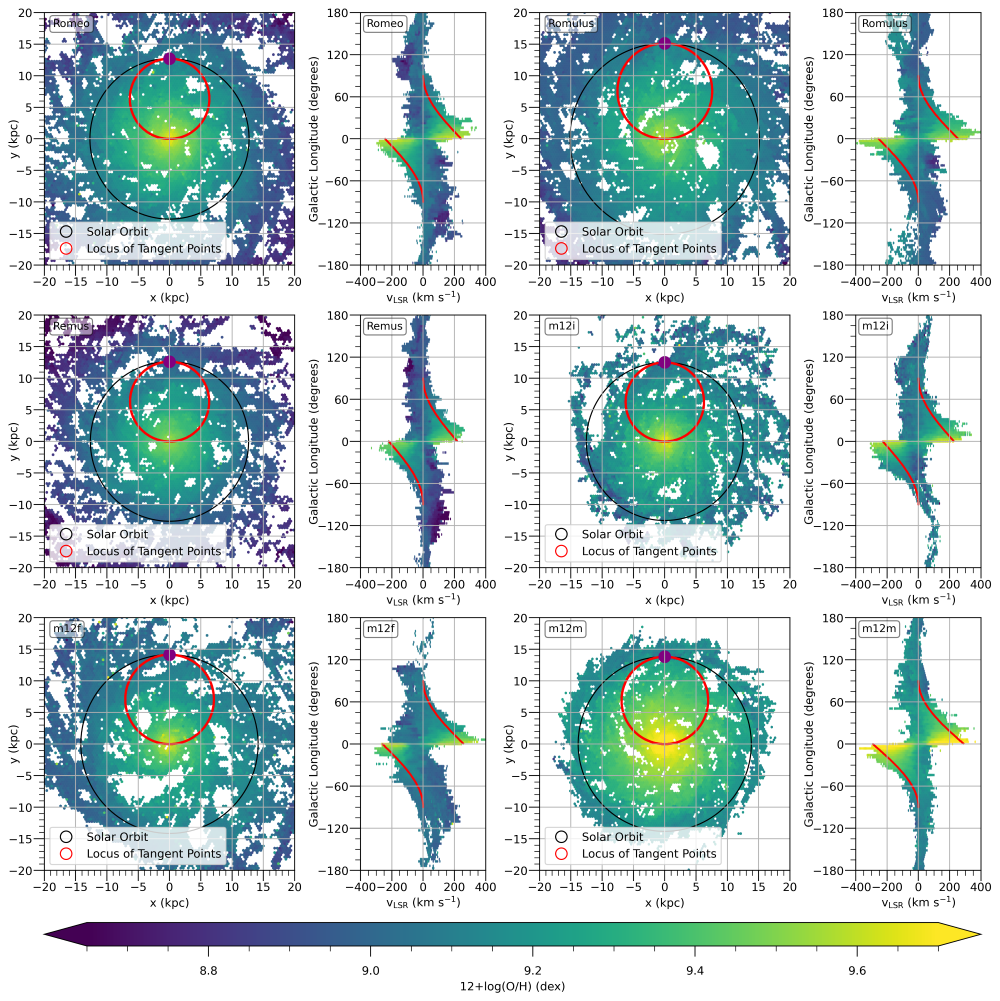}
    \caption{Face-on maps and $\ell$-$v$ diagrams for C\&D gas metallicity in all six FIRE-2 simulations in Table \ref{tab:sim_properties}, all at z=0. The colors represent the O/H abundance ratio. These six galaxies were chosen for their similarity to the MW \citep{Sanderson_2020, Wetzel_2023_FIRE2DataRelease}. In all plots, the C\&D gas is binned into hexagonal areas for easier visualization. In the face-on maps, the observer is indicated by the purple dot, the red circle is the locus of tangent points (which corresponds to the terminal velocity in the $\ell$-$v$ diagrams), and the black circle is the solar orbit of the observer. In the $\ell$-$v$ diagrams, the red line is the terminal velocity curve, which uses the mean circular velocity of all C\&D gas in that particular galaxy to construct the flat rotation curve. In each panel, the position of the observer is at galactic azimuth = 90$\degree$ with a galactocentric radius of the Sun scaled to the FIRE-2 galaxy's virial radius (see Table \ref{tab:sim_properties} for the specific $r_{\rm obs}$ values). Small-scale variations in the shape of the $\ell$-$v$ diagrams and the slope of the negative radial metallicity gradient exist between the six galaxies, but overall, all six exhibit a clear negative radial gradient in their $\ell$-$v$ diagrams.}
    \label{fig:lv_multiple_galaxies}
\end{figure}

\begin{figure}[h]
    \centering
    \includegraphics[width=1\linewidth]{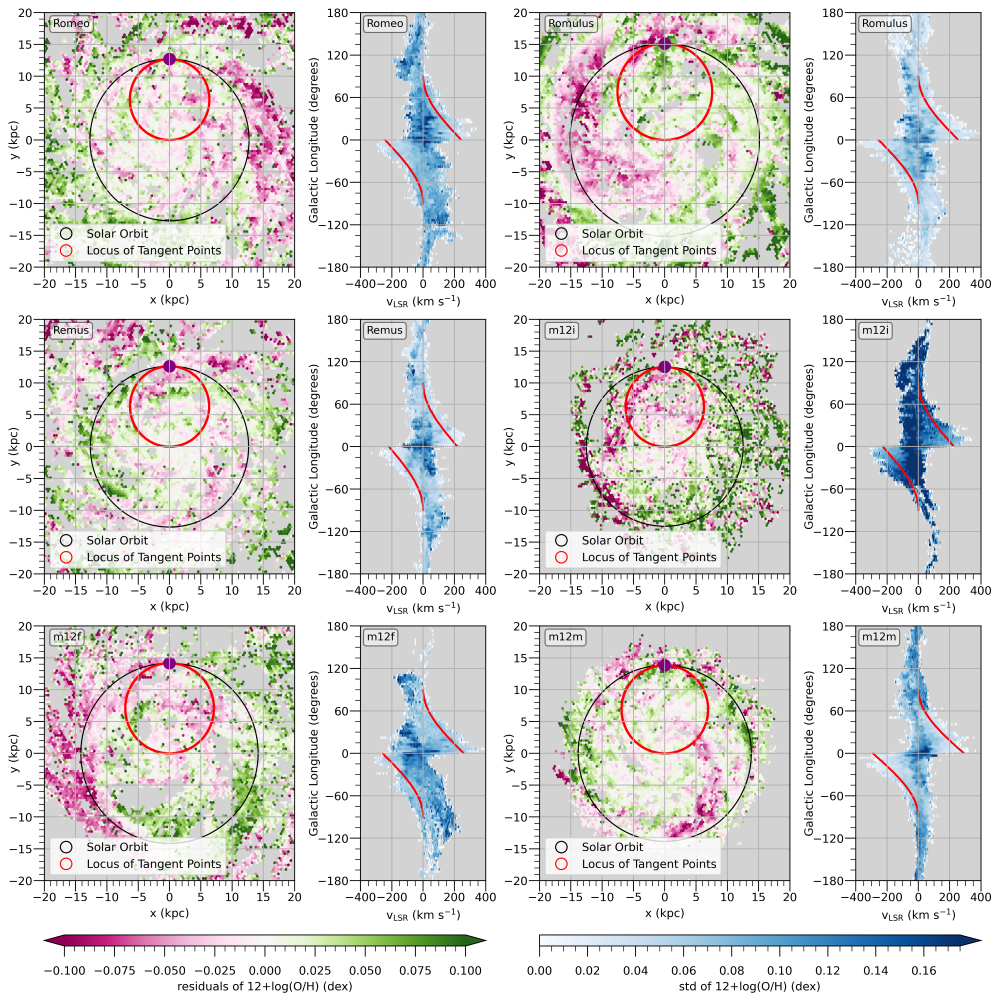}
    \caption{Face-on maps and $\ell$-$v$ diagrams for C\&D gas metallicity residual in all six FIRE-2 simulations in Table \ref{tab:sim_properties}, all at z=0. In the face-on maps, the colors represent the residual O/H abundance ratio after subtracting off the radial metallicity gradient from the total O/H abundance ratio. In the $\ell$-$v$ diagrams, the C\&D gas is binned into hexagonal areas and each bin is colored by the standard deviation of the oxygen abundance of cells in that bin. Otherwise, the figure details are the same as Figure \ref{fig:lv_multiple_galaxies}. In all six galaxies' $\ell$-$v$ diagrams, the oxygen abundance scatter is larger for bins near $v_{\rm LSR}=0$ compared to bins near the terminal velocity curve, indicating the presence of azimuthal metallicity variations.}
    \label{fig:lv_std_multiple_galaxies}
\end{figure}

By examining the $\ell$-$v$ diagram of the Romeo galaxy, we determine that both the negative radial metallicity gradient and non-axisymmetric azimuthal variations are visible in $\ell$-$v$ space. To demonstrate that $\ell$-$v$ diagrams can detect these metallicity trends in the FIRE-2 suite in general, we show $\ell$-$v$ diagrams of C\&D gas in six MW-mass galaxies colored by the total O/H abundance ratio in Figure \ref{fig:lv_multiple_galaxies} and by the standard deviation of the O/H abundance ratio in Figure \ref{fig:lv_std_multiple_galaxies}. Both the radial gradient and azimuthal variations are detected in the all six FIRE-2 simulations listed in Table \ref{tab:sim_properties}, demonstrating that using $\ell$-$v$ diagrams to study metallicity structure is widely applicable to MW-mass galaxies.

\begin{figure}[h]
    \centering
    \includegraphics[width=1\linewidth]{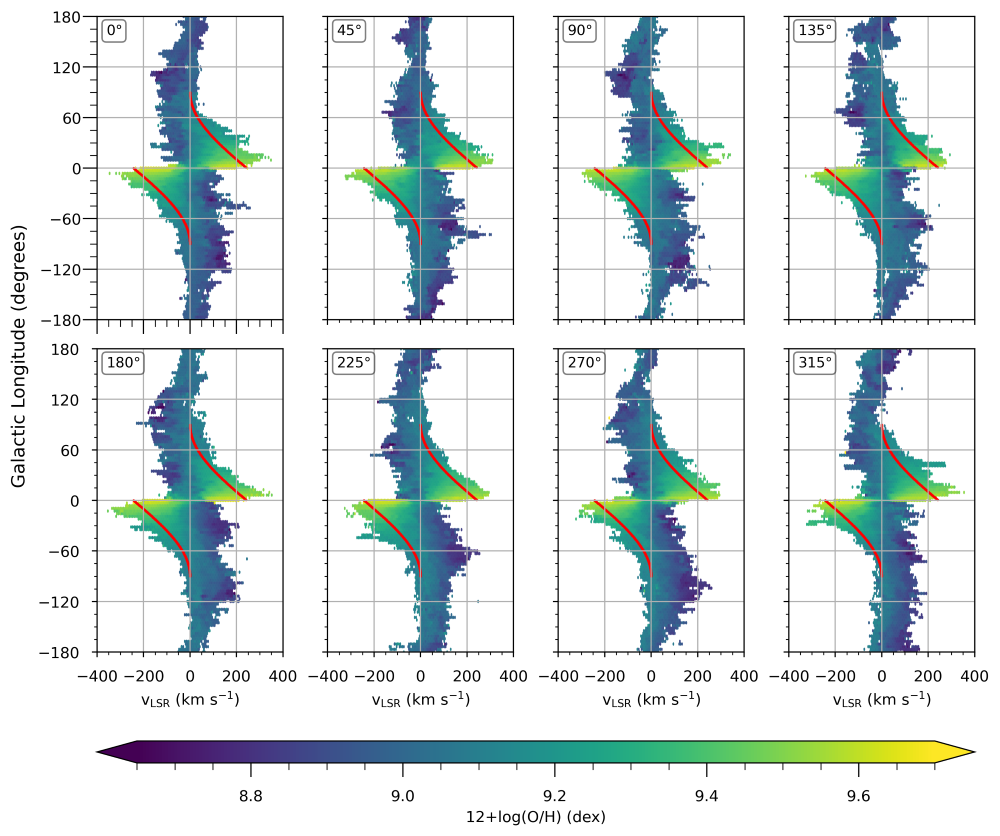}
    \caption{$\ell$-$v$ diagrams for C\&D gas metallicity in the Romeo simulation at z=0 with different observer azimuths ($\phi_{\rm obs}$). Plotted are the C\&D gas cells binned into hexagonal areas for easier visualization. Each plot is colored by oxygen abundance, and the $\phi_{\rm obs}$ is indicated in top-left corner. In the $\ell$-$v$ diagrams, the red line is the terminal velocity curve, which uses the mean circular velocity of all C\&D gas in Romeo to construct the flat rotation curve. In each panel, the observer has a galactocentric radius of 12.71 kpc. Small-scale variations in the shape of the $\ell$-$v$ diagrams exist between the different observer locations, but overall, all six positions exhibit the negative radial gradient in their $\ell$-$v$ diagrams regardless of the specific $\phi_{\rm obs}$.}
    \label{fig:lv_multiple_angles}
\end{figure}

\begin{figure}[h]
    \centering
    \includegraphics[width=1\linewidth]{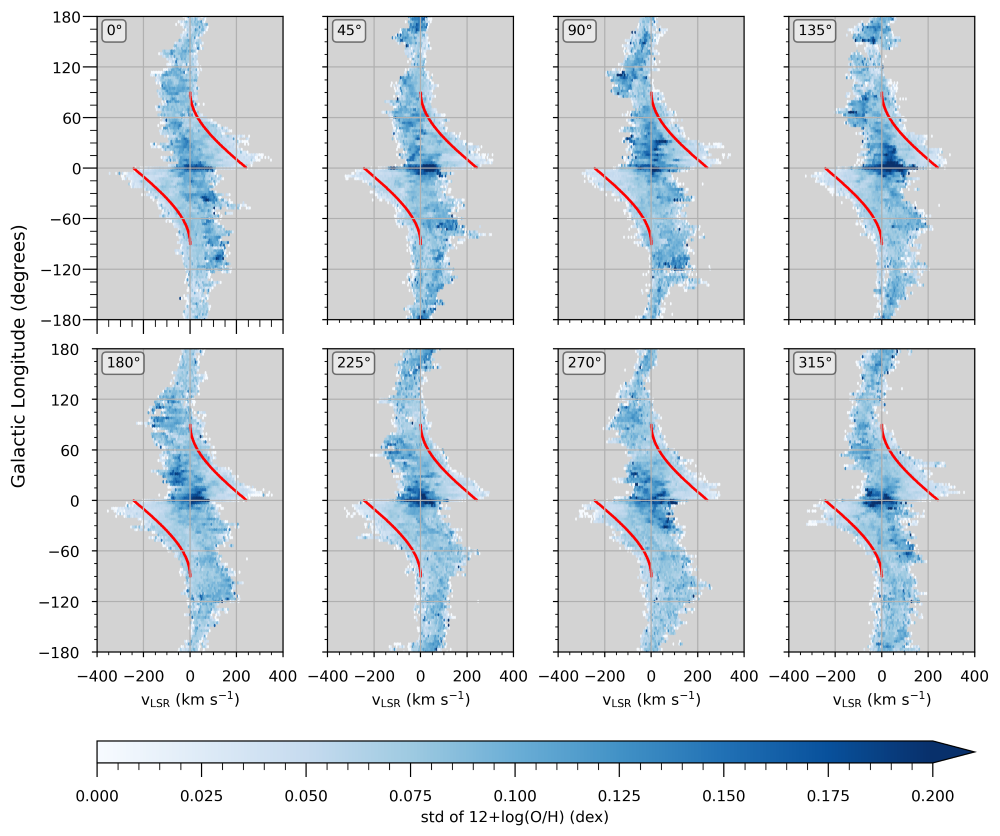}
    \caption{$\ell$-$v$ diagrams for C\&D gas metallicity variations in the Romeo simulation at z=0 with different observer azimuths ($\phi_{\rm obs}$). The C\&D gas is binned into hexagonal areas in $\ell$-$v$ space and each bin is colored by the standard deviation of the oxygen abundance of cells in that bin. The $\phi_{\rm obs}$ is indicated in the top-left corner. The red line is the terminal velocity curve, which uses the mean circular velocity of all C\&D gas in that particular galaxy to construct the flat rotation curve. In each panel, the observer has a galactocentric radius of 12.71 kpc. For all angles, the oxygen abundance scatter is larger for bins near $v_{\rm LSR}=0$ compared to bins near the terminal velocity curve, indicating the presence of azimuthal metallicity variations regardless of the specific $\phi_{\rm obs}$.}
    \label{fig:lv_std_multiple_angles}
\end{figure}

\begin{figure}
    \centering
    \includegraphics[width=1\linewidth]{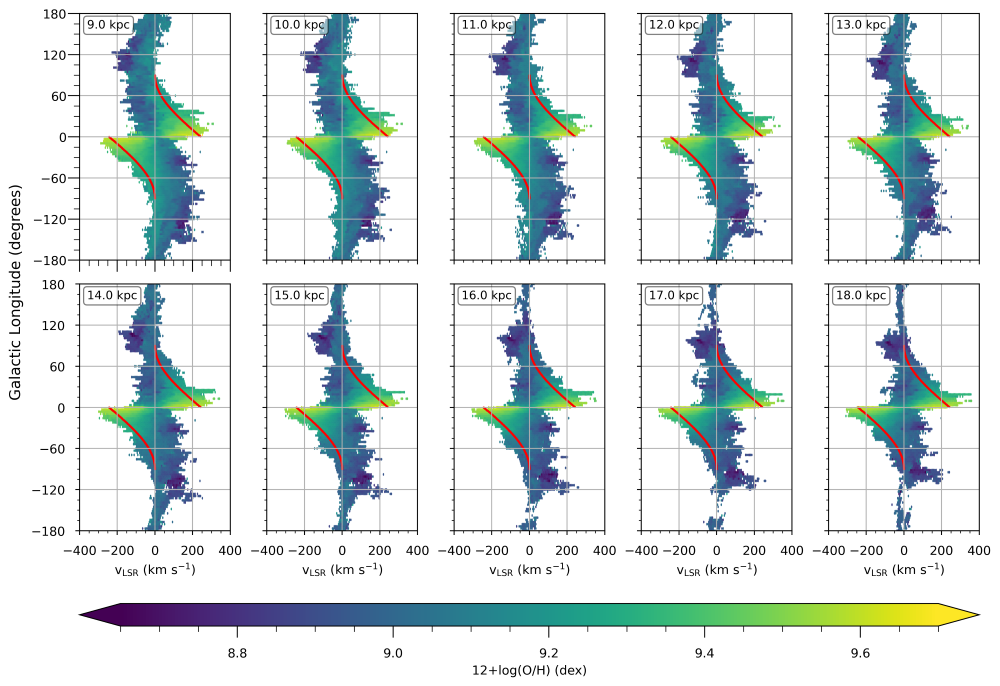}
    \caption{$\ell$-$v$ diagrams for C\&D gas metallicity in the Romeo simulation at z=0 with different observer galactocentric radii ($R_{\rm obs}$). Plotted are the C\&D gas cells binned into hexagonal areas for easier visualization. Each plot is colored by oxygen abundance, and the $R_{\rm obs}$ is indicated in the top-left corner. In the $\ell$-$v$ diagrams, the red line is the terminal velocity curve, which uses the mean circular velocity of all C\&D gas in Romeo to construct the flat rotation curve. In each panel, the observer's galactic azimuth is $\phi_{\rm obs}= 90\degree$. Small-scale variations in the shape of the $\ell$-$v$ diagrams exist between the different observer locations, but overall, all ten positions exhibit the negative radial gradient in their $\ell$-$v$ diagrams regardless of the specific $R_{\rm obs}$.}
    \label{fig:lv_multiple_rgal}
\end{figure}

\begin{figure}
    \centering
    \includegraphics[width=1\linewidth]{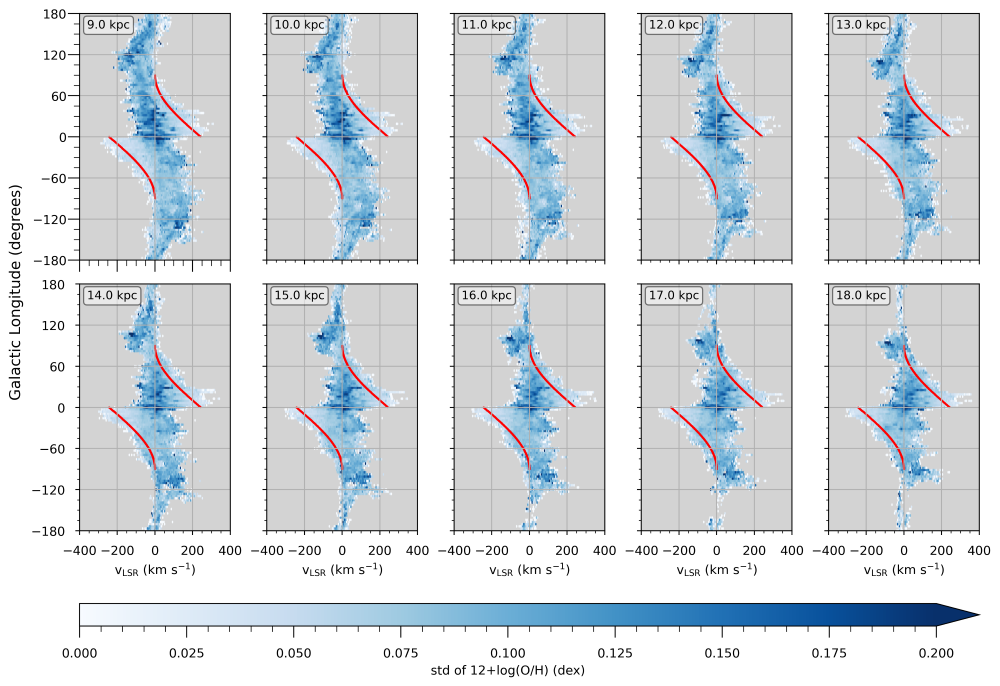}
    \caption{$\ell$-$v$ diagrams for C\&D gas metallicity variations for the Romeo simulation at z=0 with different observer galactocentric radii ($R_{\rm obs}$). The C\&D gas is binned into hexagonal areas in $\ell$-$v$ space and each bin is colored by the standard deviation of the oxygen abundance of cells in that bin. The $R_{\rm obs}$ is indicated in the top-left corner. The red line is the terminal velocity curve, which uses the mean circular velocity of all C\&D gas in that particular galaxy to construct the flat rotation curve. In each panel, the observer's galactic azimuth is $\phi_{\rm obs}= 90\degree$. For all ten plots, the oxygen abundance scatter is larger for bins near $v_{\rm LSR}=0$ compared to bins near the terminal velocity curve, indicating the presence of azimuthal metallicity variations regardless of the specific $R_{\rm obs}$.}
    \label{fig:lv_std_multiple_rgal}
\end{figure}

We also demonstrate that the detection of metallicity structure in $\ell$-$v$ space is not dependent on the location of the observer by showing the $\ell$-$v$ diagrams of C\&D gas metallicity at different observer azimuths (Figures \ref{fig:lv_multiple_angles}
 and \ref{fig:lv_std_multiple_angles}) and at different galactocentric radii (Figures \ref{fig:lv_multiple_rgal} and \ref{fig:lv_std_multiple_rgal}). In all four figures, both the radial gradient and azimuthal variations are detected, showing that the detection of metallicity structure in $\ell$-$v$ space is insensitive to the specific observer location.

\section{Mass Distribution in FIRE-2 Simulations} \label{sec:appendix_mass_dist}
\begin{figure}
    \centering
    \includegraphics[width=1\linewidth]{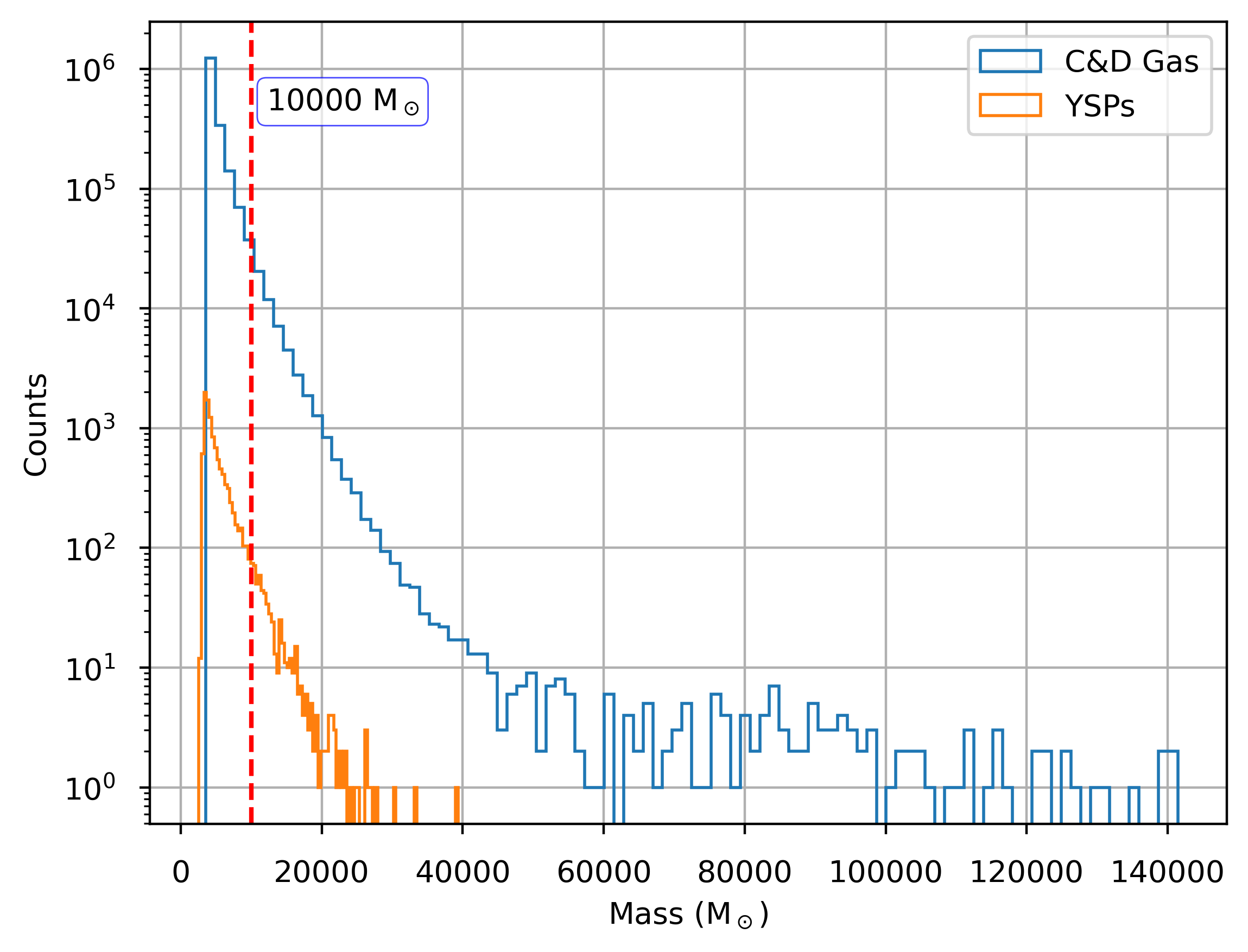}
    \caption{The mass distribution of C\&D gas (blue) and YSPs (orange) in the Romeo simulation at z=0. Note that the tick marks for the y-axis are displayed in log-scale to more clearly show the mass distributions. The dashed red line is the marker for 10000 $M_\odot$. In the C\&D gas mass distribution, $\sim$96.72\% of the gas cells have masses between 3500 $M_{\rm sun}$ and 10000 $M_{\rm sun}$, while the remaining $\sim$3.28\% have masses between 10000 $M_{\rm sun}$ and 145000 $M_{\rm sun}$. In the young stellar population mass distribution, $\sim$94.47\% of the particles have masses between 2500 $M_{\rm sun}$ and 10000 $M_{\rm sun}$, and the remaining $\sim$5.53\% have masses between 10000 $M_{\rm sun}$ and 40000 $M_{\rm sun}$.}
    \label{fig:mass_dist}
\end{figure}

One concern when creating $\ell$-$v$ diagrams of metallicity and metallicity scatter for the FIRE-2 simulations is that outlier particles/cells with extreme masses may cause spurious detections of the radial gradient and azimuthal variations in $\ell$-$v$ space. We show the mass distributions of C\&D gas and YSPs in the Romeo simulation at z=0 in Figure \ref{fig:mass_dist} as a supporting figure for this argument. In actuality, after testing this theory by comparing the mass-weighted versions for the $\ell$-$v$ diagrams of metallicity and metallicity scatter to their non-mass-weighted counterparts, we find that these outliers do not affect the detection of the radial gradient and azimuthal variations in $\ell$-$v$ space at all. The other five simulated FIRE-2 galaxies studied in this paper have similar C\&D gas and YSPs mass distributions as in Romeo. Their metallicity and metallicity scatter $\ell$-$v$ diagrams are similarly insensitive to mass outliers. For more details, see Section \ref{sec:discussion}.
\clearpage

\bibliographystyle{aasjournal}
\bibliography{refs}{}

\end{document}